\documentclass[a4paper,10pt]{article}

\usepackage{graphicx}
\usepackage[utf8x]{inputenc}
\usepackage{epstopdf}
\usepackage{graphicx}
\usepackage[normal]{subfigure}
\usepackage{overpic}
\usepackage{float}
\usepackage{parskip}
\usepackage{natbib}
\usepackage{float}
\usepackage{dblfloatfix}
\usepackage{fixltx2e}
\usepackage{color}
\usepackage{geometry}
\usepackage{amsmath}
\usepackage{amssymb}
\usepackage{array}
\usepackage{lineno}
\usepackage[american]{babel}
\usepackage{times}
\usepackage{mathtools}
\usepackage{setspace}
\usepackage{multirow}
\usepackage{rotating}

\usepackage[amssymb]{SIunits}

\title{Three-dimensional phase-field study of crack-seal microstructures -
Insights from innovative post-processing techniques}

\author{Kumar Ankit$^{1,*}$,  Michael Selzer$^{2}$,  Britta Nestler$^{1,2}$ \\ \\
$^{1}$\small\textit{IMP, Karlsruhe University of Applied Sciences,} \normalsize\textit{Moltkestr. 30, 76133 Karlsruhe, Germany} \\
$^{2}$\small\textit{IAM-ZBS, Karlsruhe Institute of Technology,} \normalsize\textit{Haid-und-Neu-Str. 7, 76131 Karlsruhe, Germany}}

\begin{document}
\date{}
\maketitle

\begin{abstract}
Numerical simulations 
of vein evolution
contribute to a better
understanding of
processes involved
in their formation and possess
the potential to
provide invaluable insights
into the rock 
deformation history and
fluid flow pathways. 
The primary aim 
of the present article 
is to investigate the influence 
of a `realistic' boundary condition, 
i.e. an algorithmically generated
`fractal' surface, on the vein 
evolution in 3-D
using a thermodynamically 
consistent approach,
while explaining the
benefits of
accounting for 
an extra dimensionality.
The 3-D simulation results 
are supplemented by innovative
numerical post-processing 
and advanced visualization techniques.
The new 
methodologies 
to measure the tracking efficiency
demonstrate the importance
of accounting the temporal evolution;
no such information is usually 
accessible
in field studies and 
notoriously difficult 
to obtain from laboratory experiments
as well.
The grain growth statistics 
obtained by numerically
post-processing the
3-D computational 
microstructures explain the 
pinning mechanism 
which leads to 
arrest of grain boundaries/multi-junctions
by crack peaks,
thereby, enhancing the
tracking behavior.

\end{abstract}

 \let\thefootnote\relax\footnotetext{$^{*}$Corresponding author: Kumar Ankit, Email: \textit{kumar.ankit@hs-karlsruhe.de}}

\section{Introduction}

Correct interpretation of vein microstructure
demands careful study of thin sections
and appropriate bridging of results obtained
from different length scales.
The field based studies as well as
laboratory experiments are the two different means 
by which geologists try to establish 
structure-property co-relationship.
Unless one performs arduous in-situ
laboratory experiments, the time-based
evolution information stays inaccessible.
Moreover, it is difficult to decompose the
effect of different processes that might have
acted in sequence or simultaneously. 
\citet{Piazolo:2010uq} summarize the limitations
of field based studies and laboratory 
experiments while highlighting the 
general capability
of numerical simulations to serve as
a viable alternative. 

\begin{figure}[H]
\centering
\subfigure[]{\includegraphics[scale=0.22]{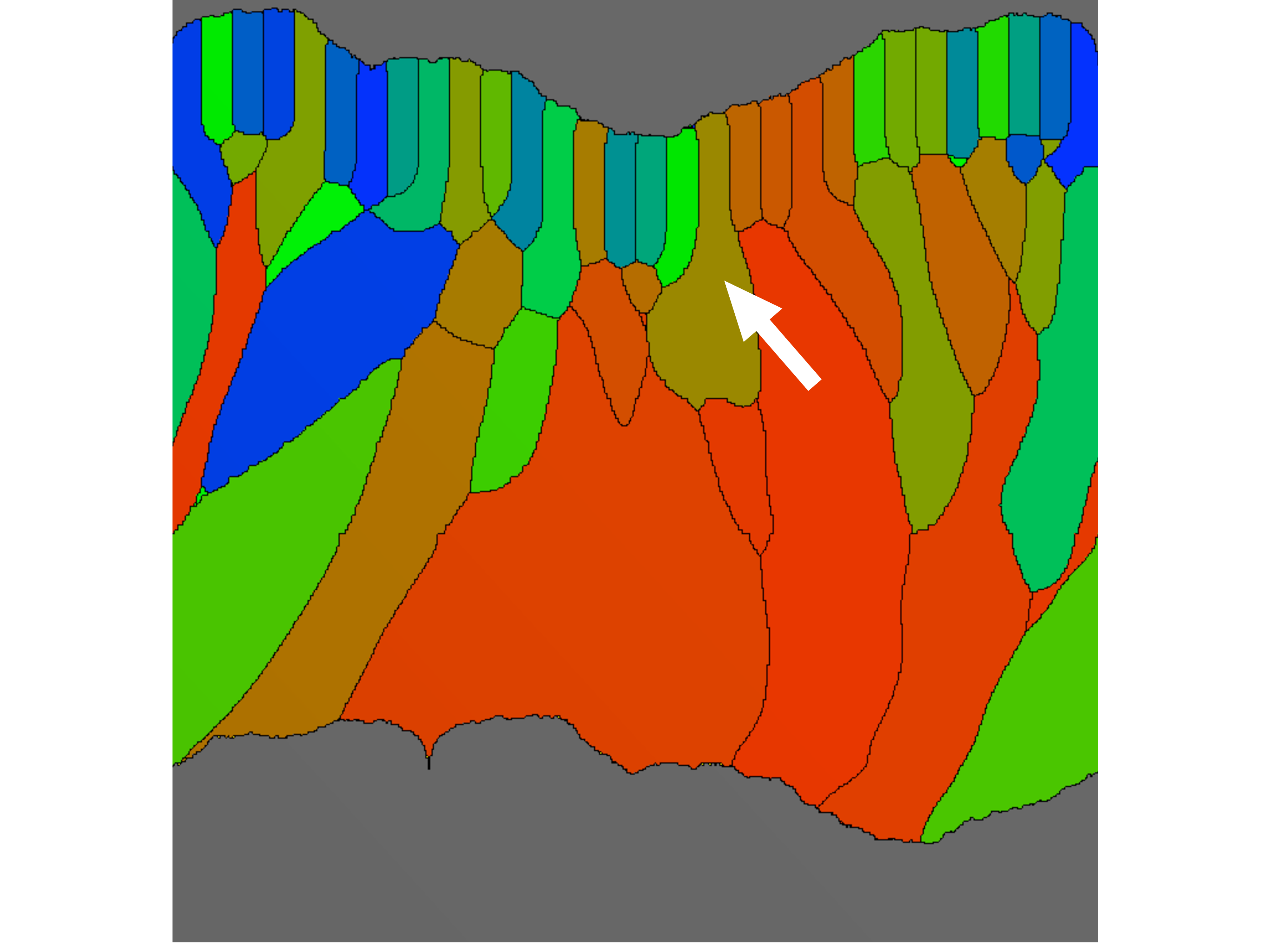}\label{fig:thin_section}}\hspace{1.5 cm}
\subfigure[]{\includegraphics[scale=0.235]{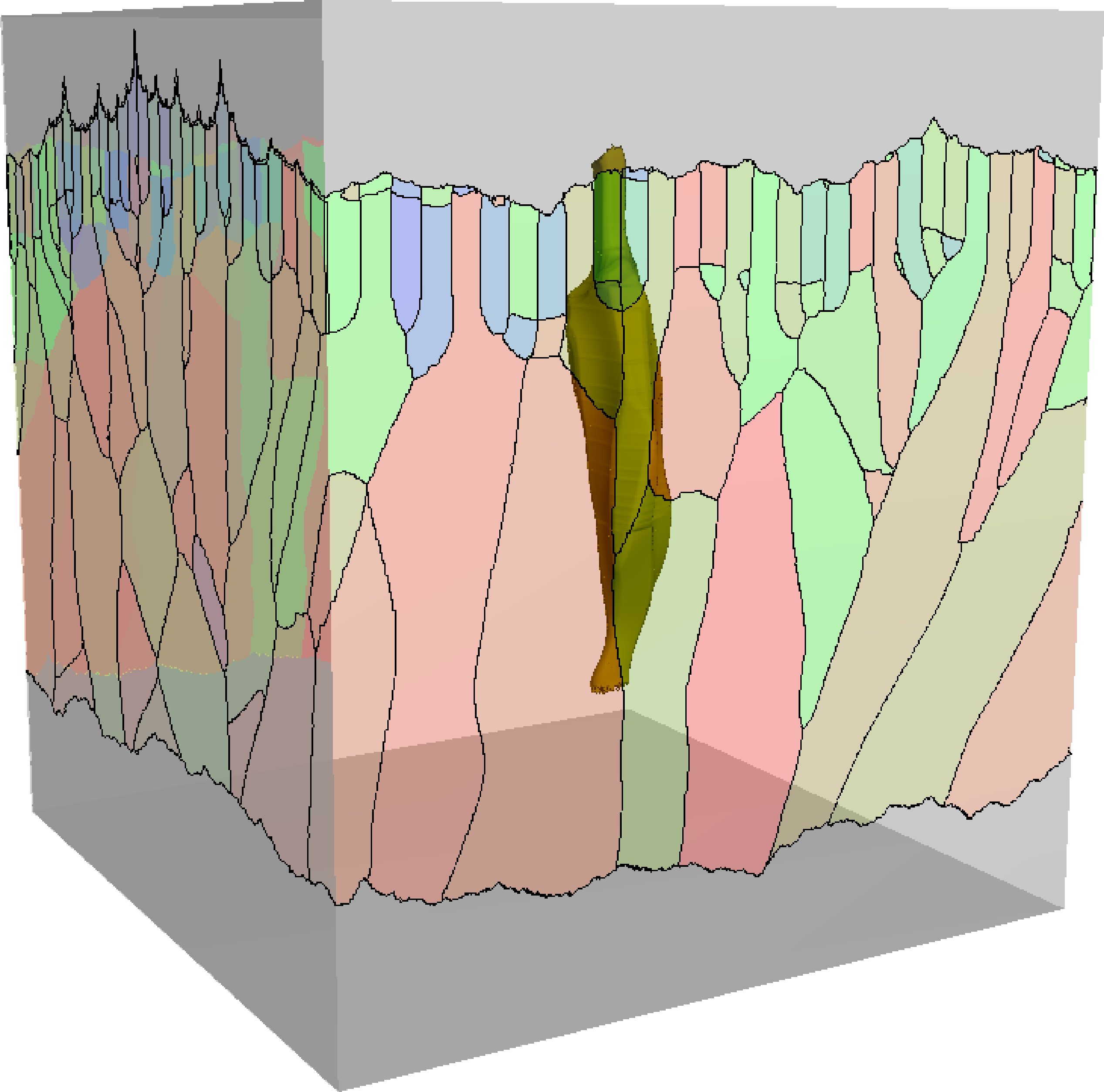}\label{fig:transparent_cube}}
\caption{Three dimensional phase-field simulation of the crack-seal microstructure.
The direction of crack opening is along a 
quarter circle towards the left in the plane of 
computational thin-section.
(a) Thin section cut of a 3-D computational 
microstructure showing the consumption 
of a grain as indicated by the white arrow.
(b) The 3-D computational microstructure 
exposes that the grain was not consumed;
rather it evolved along a different plane. 
Transparencyof the surrounding grains 
enable the visualization of the 
tracking vein  inside the 
numerical domain.
}
\label{fig:err_interpretation} 
\end{figure}

The need for numerical 
simulations to study microstructural
evolution in veins has lead to development
of computer programs, based on front-tracking
approaches, like \textit{Vein Growth}
\citep{Bons:2001ve} and FACET \citep{Zhang:2002fk}
in the past and more recently, 
\textit{Elle} \citep{Jessell:2001kx,Bons:2008ys}. 
A major limitation of
such front-tracking methods
is that numerical studies require 
huge effort to be extended to 3-D. 
The idea of comparing the 2-D
numerical results with
thin sections of 3-D natural samples
 is far-fetched, primarily because of 
 an added degree of freedom for the 
 evolving crystals (in 3-D) which leads to 
 erroneous interpretation.
While \citet{Bons:2001ve} provide hints
concerning deviations when thin sections are
directly compared with 2-D simulation results,
the actual differences remain unclear due to lack of 
any 3-D numerical results. 
It is noteworthy,
that a complete understanding
of the vein growth mechanism 
cannot be achieved by merely
studying thin sections as it 
reveals only a part of the whole story.
The partial story could often be misleading,
since what looks to be a crystal not tracking the 
crack-opening trajectory in 2-D microstructure
may turn out to be exactly opposite
when studied in 3-D.
In the context of
vein growth, we emphasize 
that the grain formation
process is generically of
3-D nature and
can be interpreted
in a physically sufficient manner
by methods capable of
capturing the growth
characteristics and dynamics
in full 3-D space.
The phase-field
simulation example shown in
figure \ref{fig:err_interpretation}
bear testimony to our
assertion.

The last two decades have seen 
the emergence of phase-field 
method as a versatile and 
popular tool for numerical
simulations of crystal growth 
and in general,
for a variety of other moving 
boundary problems;
prominent application areas being
solidification, solid-state phase
transformations like spinodal
decomposition and crack propagation. 
The most significant computational 
advantage of a phase-field model 
is that explicit tracking of the interface 
is unnecessary. In simpler words, in a 
phase-field simulation, freely moving 
interfaces between different phases
or crystals (grain orientations) do 
not appear as geometric boundaries, i.e. 
places at which boundary conditions 
have to be applied explicitly. 
Instead, all the information about 
the location of the phase boundaries 
is implicitly contained in the phase-field variable, 
which obeys a partial differential equation 
and solved within the entire 
computational domain.
Different thermodynamic driving forces 
for topological changes, such as chemical 
bulk free energy, interfacial energy, elastic 
strain energy and different transport processes, 
such as mass diffusion and advection, 
can be coupled, thereby facilitating 
the comprehensive studies
of the transformation phenomena.

\citet{HUBERT:2009fu} discuss the limitations
of \textit{Vein Growth} and FACET
(geometric restrictions and extension
to 3-D) and
introduce a phase-field
model to study crystal 
growth in veins, which uses a
non-faceted anisotropy for
the interfacial energy. 
However, it is well known 
from literature 
\citep{Taylor:1998kq}
that smooth continuous functions
cannot be used to simulate 
crystals with flat facets 
and sharp corners.
\citet{Ankit:2013qf} show the
importance of adopting a 
general thermodynamically
consistent approach in modeling 
the evolution of vein microstructures
by considering faceted-type 
anisotropy formulations
of the interfacial 
energy function to simulate
crystals with flat facets and
sharp corners.
Various boundary conditions
and parameters
which influence the
crystal growth mechanism 
in veins,
especially the grain boundary
tracking behavior, can be
successfully investigated
using the phase-field method.
Further, the reproducibility
of previous simulation results
(from front-tracking models)
as well as chief advantages
of adopting the novel 
multiphase-field model, 
such as 3-D numerical studies for 
crystal of any shape, 
large-scale simulations
with many thousand grains
and provision to 
implement transport, 
is highlighted.

In the current article, we 
advance the 
3-D numerical study of \citet{Ankit:2013qf}
to study the influence of a
realistic representation
of the crack wall by a 
boundary condition
of fractal shape,
on crystal growth in crack-sealing
veins. By employing
advanced visualization
and innovative post-processing
techniques, new methodologies
to calculate 'general 
tracking efficiency' 
for a more complex 
motion of grain boundaries
is proposed. The present work 
highlights the importance of 
accounting the time evolution 
rather than calculating tracking 
efficiency solely based on final
grain boundary morphology.
Further, grain statistics such as 
temporal evolution of the number of
tracking veins and the corresponding 
orientation and size distribution is 
obtained from the present large scale
3-D phase-field simulations with an
aim to relate
the shift in 
growth mechanism
as a function of 
crack-opening rate,
which is missing in 
previous numerical studies.

\section{Methods}
\subsection{Phase-field model}
The foundation of multiphase-field method utilized to address crystal growth problem is 
realized by a Helmholtz free energy functional formulated as
\begin{equation}
{\cal F}\left(\boldsymbol{\phi}\right)=\int_{\Omega}\left(f\left({\boldsymbol{\phi}}\right) +\varepsilon 
a\left({\boldsymbol{\phi}},\nabla{\boldsymbol{\phi}}\right) + \frac{1}{\varepsilon}
w\left({\boldsymbol{\phi}}\right)\right)dx,
\label{eq:free_energy_func}
\end{equation}
where $f\left(\boldsymbol{\phi}\right)$
 is the bulk free energy density,
$\varepsilon$ is the small length scale parameter related to the
interface width, $a\left(\boldsymbol{\phi},\nabla\boldsymbol{\phi}\right)$ is the gradient
type and $w\left(\boldsymbol{\phi}\right)$, a potential type energy density.
The phase-field parameter $\boldsymbol{\phi}\left(\vec{x},t\right) =
\left(\phi_{1}\left(\vec{x},t\right)\cdots\phi_{N}\left(\vec{x},t\right)\right)$
describes the location of `N' crystals with different
orientation in space and time.
The evolution equations for the phase-field
vector components are 
described by variational derivative of
the free energy functional 
which ensures energy and mass conservation
as well as an increase of total entropy.
\begin{equation}
\tau\varepsilon\dfrac{\partial\phi_{\alpha}}{\partial t}=
\varepsilon\left(\nabla\cdot a_{,\nabla\phi_{\alpha}}\left(\boldsymbol{\phi},\nabla\boldsymbol{\phi}\right)-
a_{,\phi_{\alpha}}\left(\boldsymbol{\phi},\nabla\boldsymbol{\phi}\right)\right)-
\dfrac{1}{\varepsilon}w_{,\phi_{\alpha}}\left(\boldsymbol{\phi}\right)-
f_{,\phi_{\alpha}}\left(\boldsymbol{\phi}\right)-\lambda
\label{eq:evolution_eqn}
\end{equation}
The symbol $\tau$ is the kinetic coefficient and comma separated sub-indices represent
derivatives with respect to $\phi_{\alpha}$ and 
gradient components $\dfrac{\partial\phi_{\alpha}}{\partial\chi_{i}}$.
The lagrange multiplier $\lambda$ guarantees the
summation constraint 
$\left(\displaystyle\sum_{\alpha=1}^{N}\;\phi_{\alpha}=1\right)$.
The multiphase-field model 
equations and numerical 
implementation have been 
previously discussed in 
detail \citep{Nestler:2005ye, Stinner:2004uq}.
The phase-field evolution
equation \ref{eq:evolution_eqn}
is non-dimensionalised
to ensure numerical
accuracy during the computation.
A brief account of the 
non-dimensionalisation
procedure is provided
by \citet{Wendler:2009fk}.

\begin{figure}[H]
\centering
\subfigure[]{\includegraphics[scale=0.08]{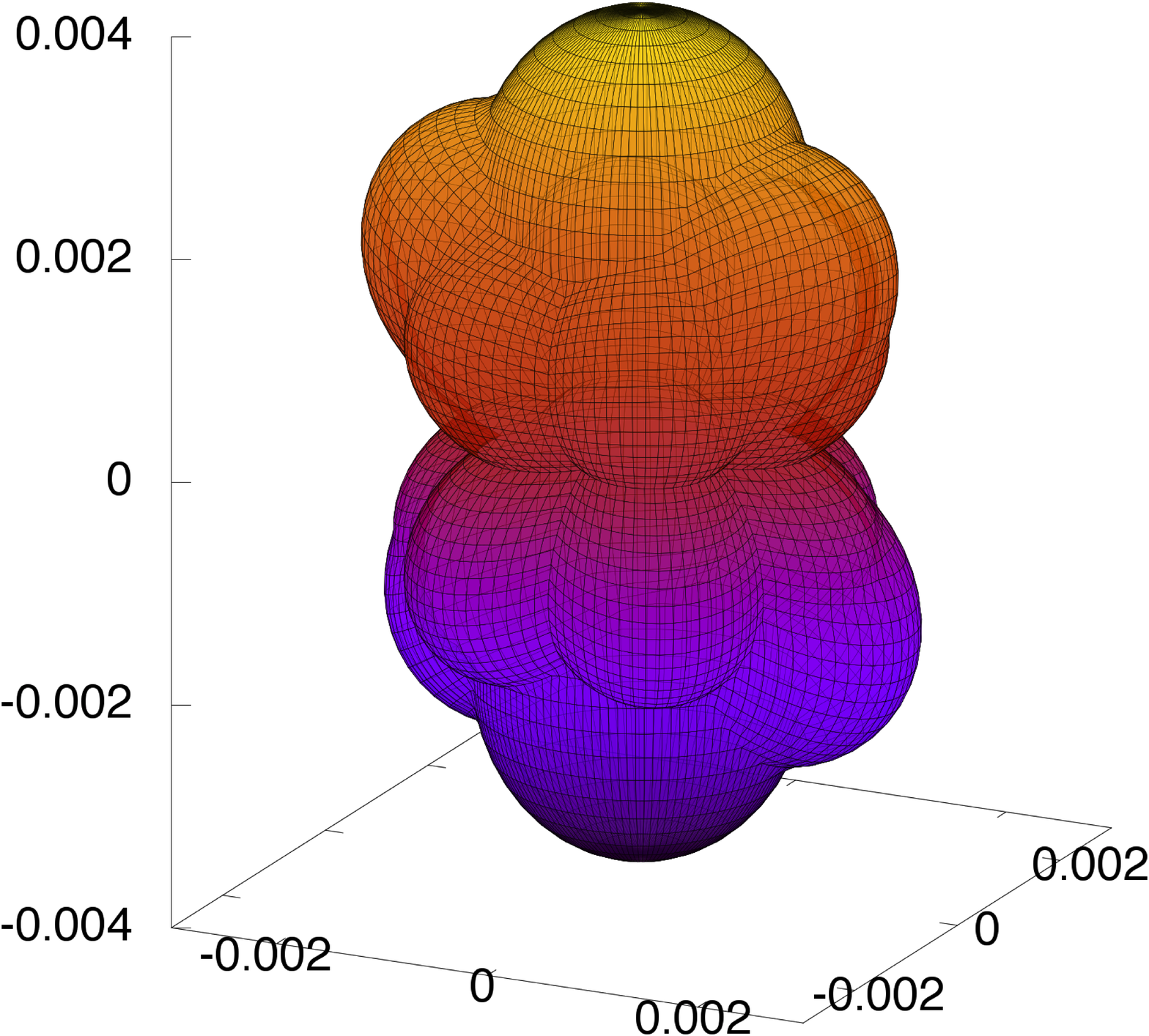}\label{fig:quartz_gammaplot}}
\subfigure[]{\includegraphics[scale=0.11]{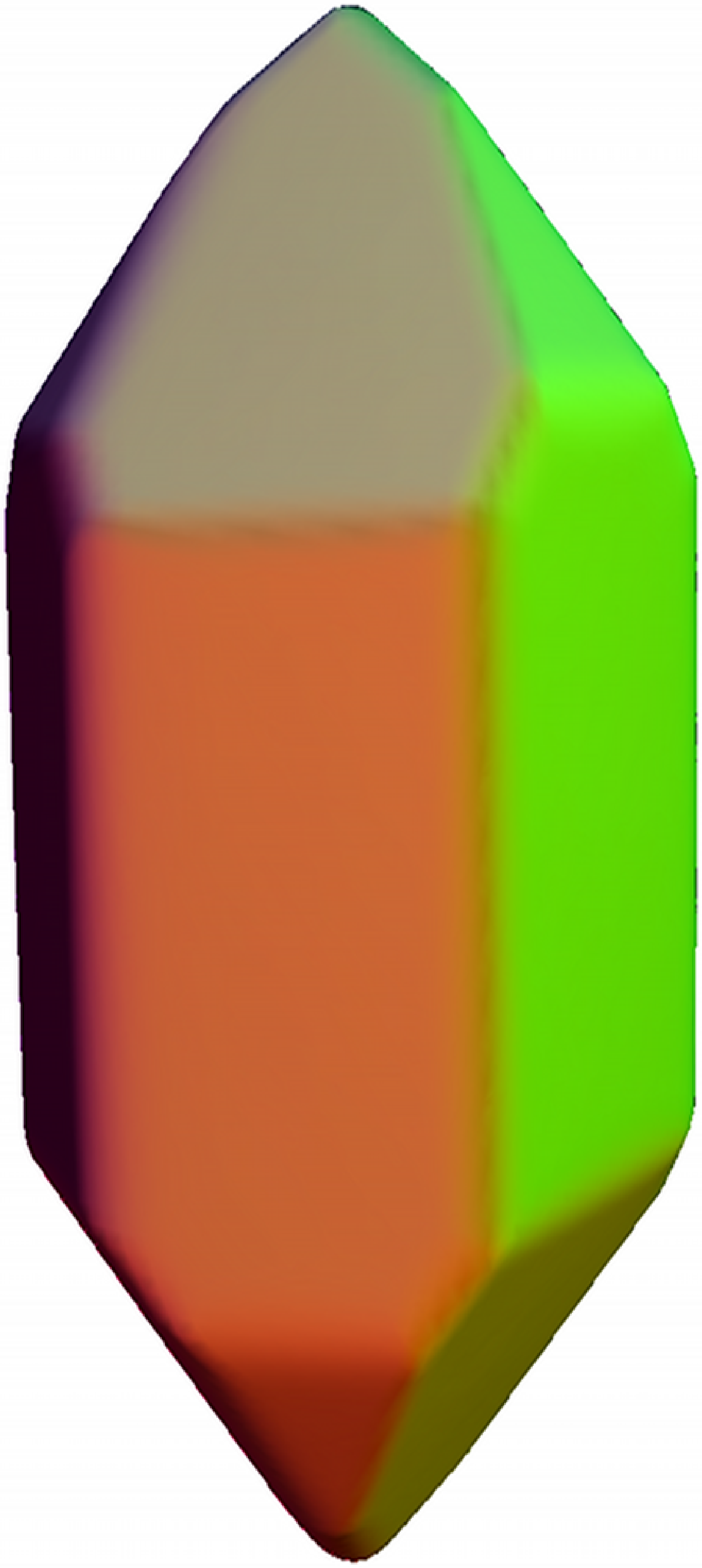}\label{fig:quartz_wulff}}
\subfigure[]{\includegraphics[scale=0.19]{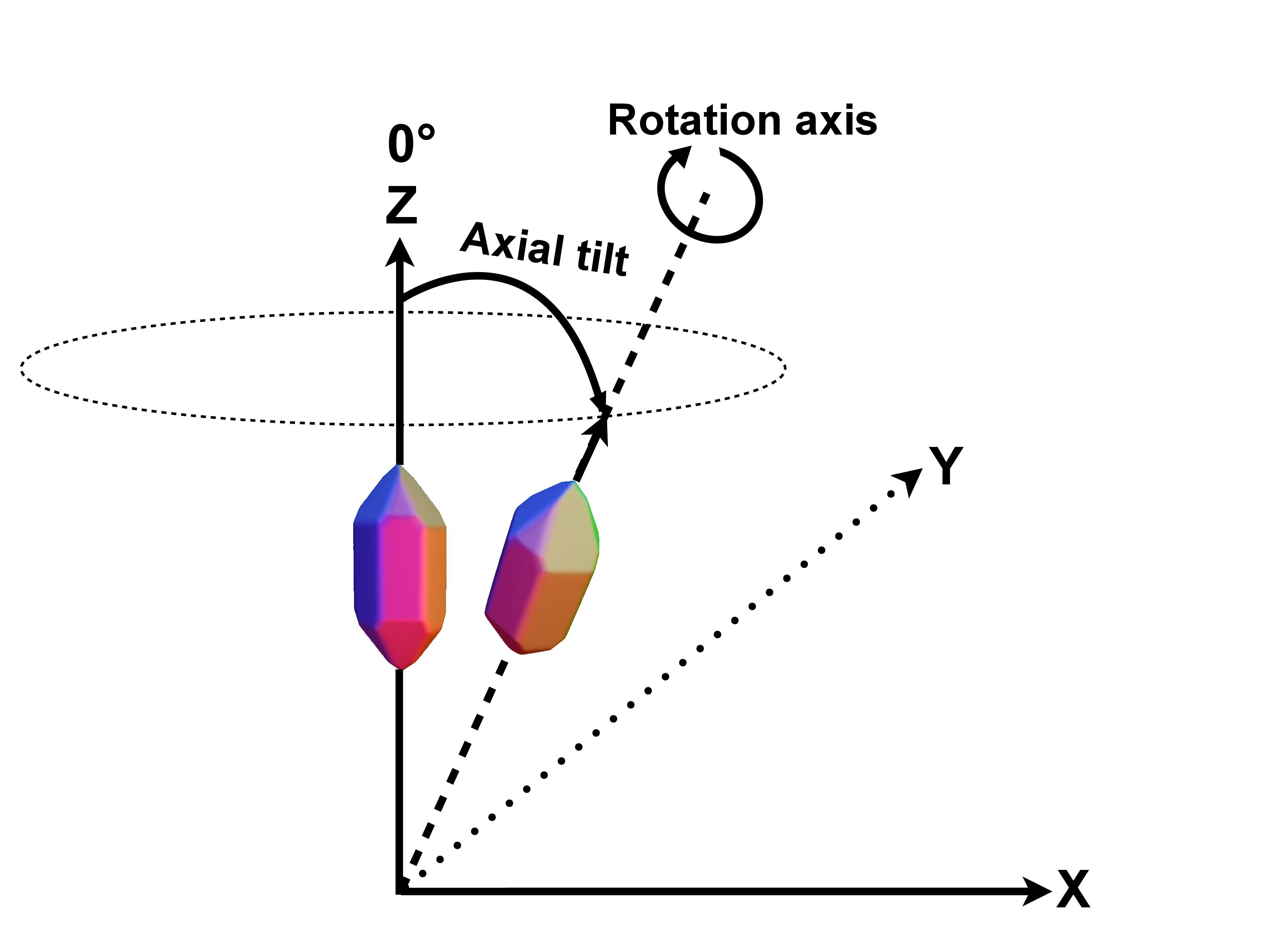}\label{fig:orient_definition}}
\caption{(a) Polar plot of the interfacial energy for
the symmetry of a quartz crystal. 
(b) Equilibrium quartz shape obtained from phase-field simulation.
(c) The definition of crystal orientation in 3-D.
The colors in figure \ref{fig:quartz_wulff} and 
\ref{fig:orient_definition} differentiate among crystal facets.}
\end{figure}

In the present work,
we adopt the approach
of \citet{Ankit:2013qf}
to construct the interfacial
energy function 
$a\left(\phi,\nabla\phi\right)$ for a 
geologically relevant mineral
`quartz'. The polar-plot
of interfacial energy
and the corresponding equilibrium
crystal shape is
shown in figures 
\ref{fig:quartz_gammaplot}
and \ref{fig:quartz_wulff}
respectively.
The equlibrium crystal
shape represents
an idealized quartz
crystal consisting 
of $\left\lbrace10\bar{1}0\right\rbrace$,
$\left\lbrace10\bar{1}1\right\rbrace$ 
and $\left\lbrace01\bar{1}1\right\rbrace$
facets. An implicit assumption
of the numerical simulations 
of vein growth process,
presented 
in the later sections
of this article
is that the smaller crystal
facets ($\left\lbrace2\bar{1}\bar{1}1\right\rbrace$
and $\left\lbrace6\bar{1}\bar{5}1\right\rbrace$) 
possess higher growth rate 
and hence, do not influence  the
final polycrystalline morphology.
Therefore,  
such faster growing and 
smaller facets
are ignored in the
present numerical studies. 
However, it is to be 
noted that
it is possible to simulate
such facets, provided
the polar-plot 
of interfacial energy (cusps)
accounts for them.
Hence, the assumption made
in present work should be
interpreted as shape
simplification
and \textit{not} as a limitation
of the simulation algorithm. 


\subsection{Numerical aspects}

\begin{figure}[H]
\centering
\subfigure[]{\includegraphics[scale=0.19]{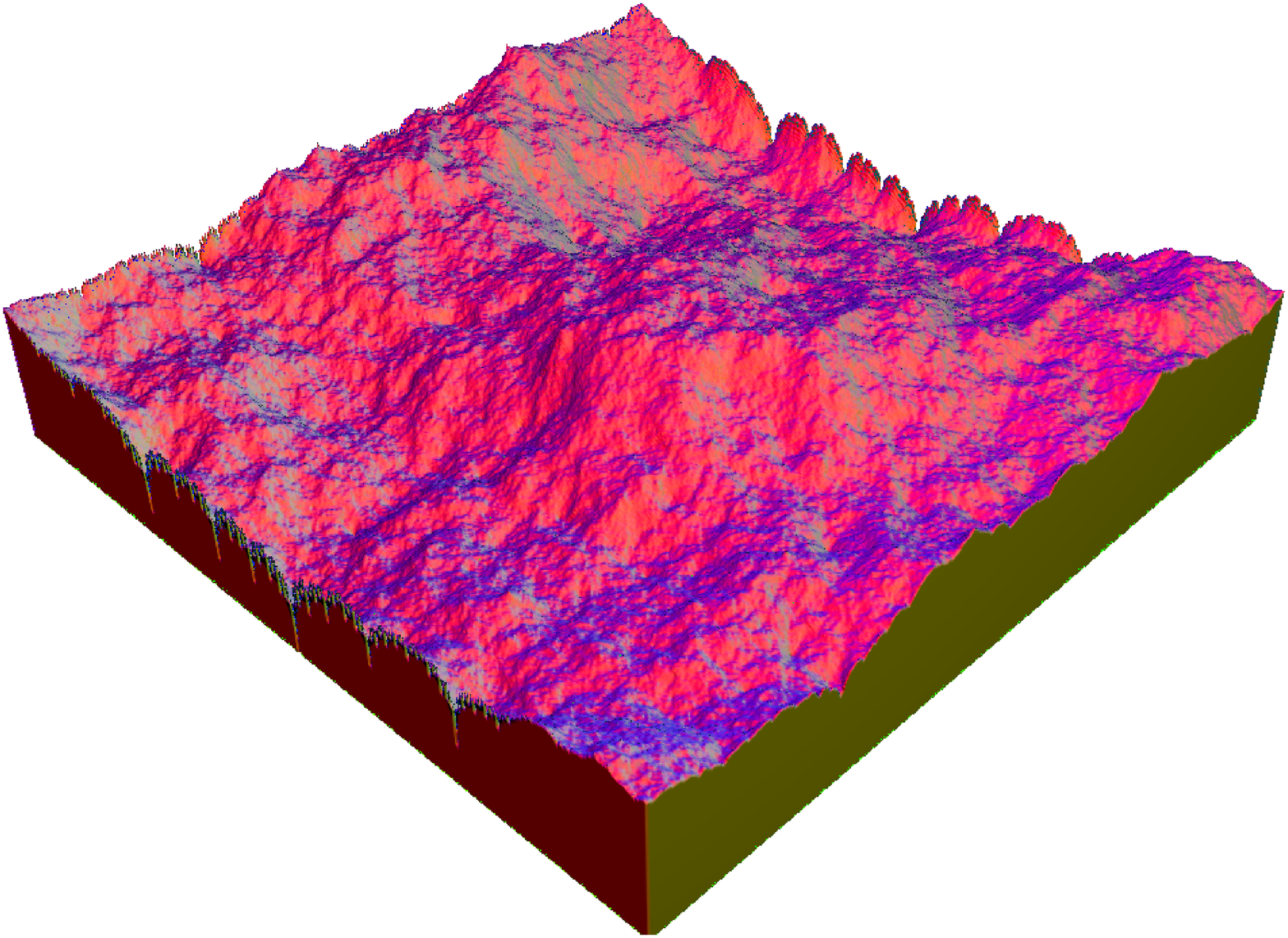}\label{fig:fractal_surface}}\qquad
\subfigure[]{\includegraphics[scale=1.9]{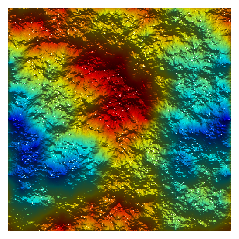}\label{fig:map_view_terrain}}
\subfigure{\includegraphics[scale=0.22]{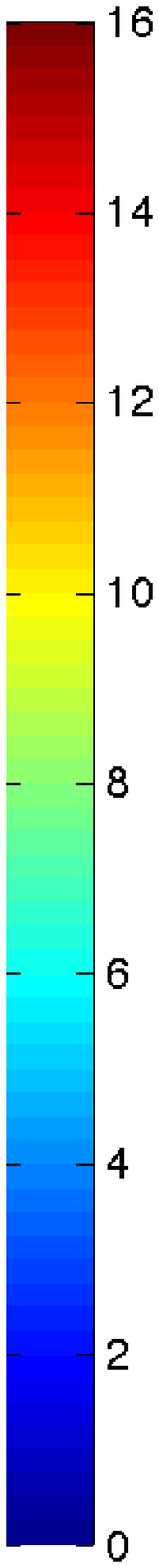}\label{fig:colormap}}
\caption{(a) Fractal surface (generated by a well-known diamond-square algorithm) 
used for 3D crack-sealing simulations. (b) Height-map of the generated surface.} \label{fig:gen_crack_surface}
\end{figure}

\begin{figure}
\centering
\includegraphics[scale=0.4]{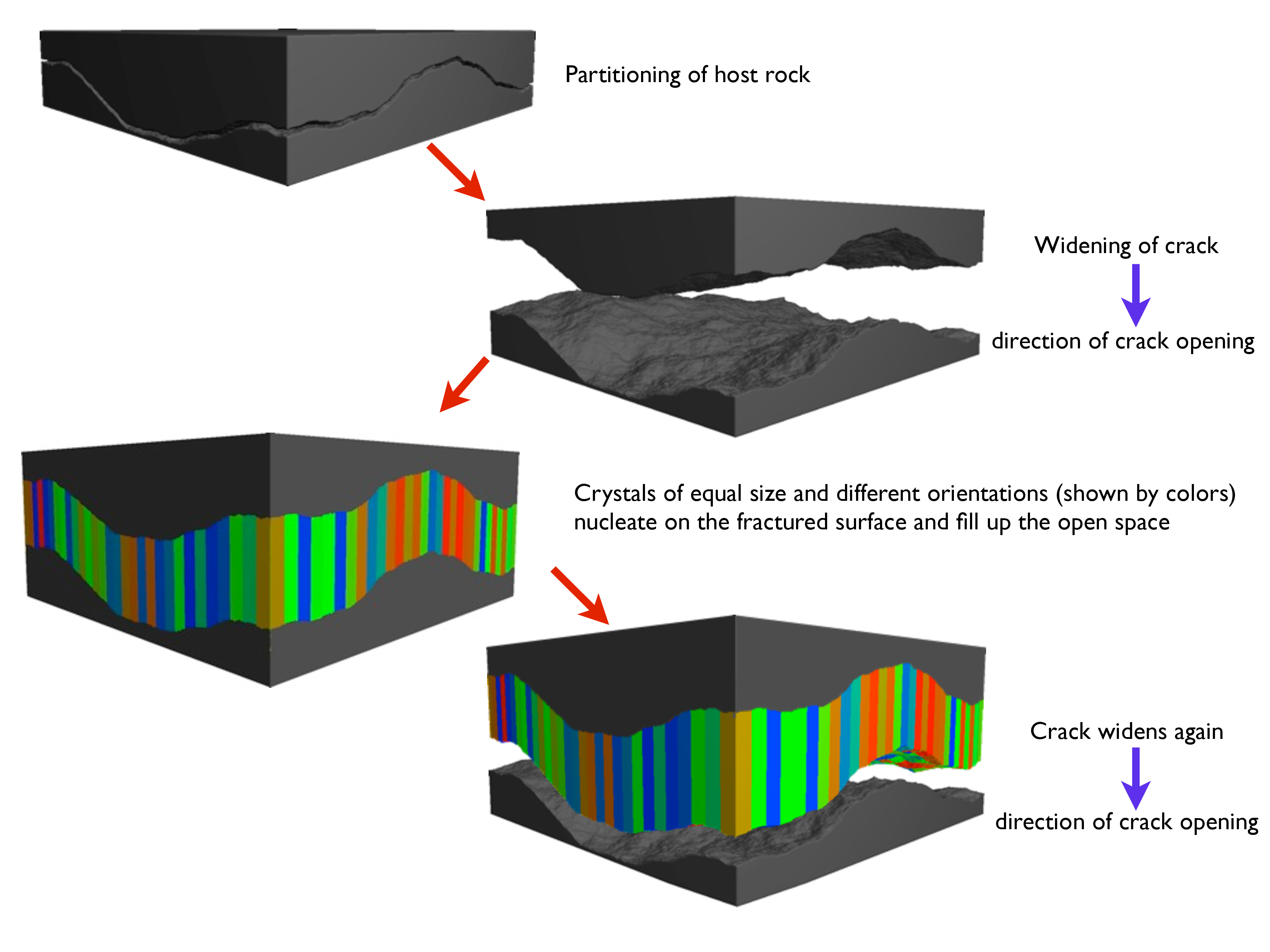}
\caption{Sequence of numerical pre-processing adopted to 
obtain a homogeneous overlay (of same size) of crystal nuclei on the algorithmically
generated fractal surface. The final domain appearing in the above sequence
is used as the initial condition for phase-field simulations.}
\label{fig:initial_condition}
\end{figure}

With an objective to numerically simulate the crack-sealing process
and to characterize the resulting microstructure, we choose quartz
crystals as a representative vein forming material. An algorithmically
generated (diamond-square algorithm) fractal surface is utilized
to model a three dimensional crack surface (boundary condition) for phase-field 
simulations as shown in figure \ref{fig:gen_crack_surface}. 
The progressive splitting of host-rock and crystal precipitation
in the open space is algorithmically replicated once by numerical pre-processing
to obtain the initial condition for simulation as shown in  figure \ref{fig:initial_condition}. 
The preprocessing algorithm adopted is as follows:
\begin{itemize}
\item The fractal topology of the 
lower crack surface is 
generated by a C program
to implement the 
 well-known diamond-square
algorithm \citep{Miller:1986qf}. 
The height-map of
upper crack surface is obtained
by subtracting the respective heights
(for lower crack-surface)
from the total height of the parent rock
in consideration.The two complementary fractal surfaces are stationed 
over each other with a minor clearance which represents the 
fracture in host rock.
\item The space between the
upper and lower surface is 
increased (by 5 grid points 
in the splitting direction) 
which represents
the first crack-opening event.
\item Crystals of equal size 
with different orientation in space 
are initially laid on the lower crack surface.
At this point, it is important to note that
the size of the crystal nuclei needs to 
be equal in order to negate 
the advent of size-effects
in the following vein growth 
simulations. The
numerical pre-processing 
technique in order to 
rule out any such possibilities
involves the following sub-steps: 
Cuboid crystal nuclei 
(different colors represent an
axial tilt defined in the previous section) 
are generated separately
and merged 
with the parent numerical domain 
containing the cleft (a boundary condition) 
such that the
former can be over-written. 
The resulting domain can be described as a perfectly 
sealed microstructure.
\item The lower crack surface is shifted downwards again to create a small
space between the wall and crystal front. Thus, we obtain the initial numerical 
domain to start the phase-field simulations.

\end{itemize}

We simulate the unitaxial opening and sealing 
of cracks to investigate the formation
of 3-D crack-sealing microstructure
using the initial condition generated
in figure \ref{fig:initial_condition},  
and thereby, characterize the 
tracking behavior of grain 
boundaries and multi-junctions.
The following discussion
will focus on the 
two 3-D simulation test cases with the same initial condition 
but different crack-opening rate, as described in table 
\ref{tab:param_table}.
\begin{table}
\caption{Table showing the
choice of 
crack-opening rate for
simulations A and B 
($\Delta x=\Delta y=\Delta z$ and
$\Delta t=0.12$).
}
\label{tab:param_table}      
\centering
\begin{tabular}{c c c c}
\hline
Simulation & Time between successive & Opening increment in  & Trajectory \\
                 & opening         & vertical direction        & \\
\hline
A & 2$\Delta t$ & 3$\Delta x$ & Quarter arc \\
B & 2$\Delta t$  & 8$\Delta x$  & Quarter arc\\
\noalign{\smallskip}\hline
\end{tabular}
\end{table}
The phase-field evolution equation 
\ref{eq:evolution_eqn} is solved 
numerically using an 
explicit forward Euler scheme. 
The spatial derivatives of the right 
hand side equation are discretized 
using a second order accurate 
scheme with a combination of 
forward and backward finite differences. 
The implementation of a locally reduced 
order parameter optimization employs a
dynamic listing of a limited number
of  locally existing grains
and enable to reduce 
computation time so that
large scale simulations in 3-D 
become feasible.
The phase-field simulations
are performed on Linux
high-performance computation
clusters using a 
C program with parallel
algorithms for domain
decomposition to
distribute the computing 
task on different nodes.

As a result of 
wall rock opening
along the predetermined
opening trajectory, the
crack-aperture increases during
simulation run-time.
This adversely affects
the computational efficiency
as the size of simulation
domain increases in the
splitting direction. 
In order to avoid such 
complications, the 
simulation is carried 
out in a 
moving frame 
(also known as shifting-box simulation). 
In the present simulations, the domain is shifted 
in the growth direction (downwards) by 
 adding a row of grid-point at the top of domain
 and discarding off a row of grid-points at the 
 bottom, every time the advancing crystal
growth front fills up 10\% of the simulation box.
The final domain is obtained by aggregating back
the discarded rows of pixels as described by 
\citet{Ankit:2013fk} for a different material system.
Further, we ensure that the advancing 
crystal-rock interface always stay within 
the boundaries of shifting-box for every simulation
time-step.

\section{Simulation results}
\subsection{Calculation of tracking efficiency}

\begin{figure}[!htbp]
\centering
\subfigure[]{\includegraphics[scale=0.15]{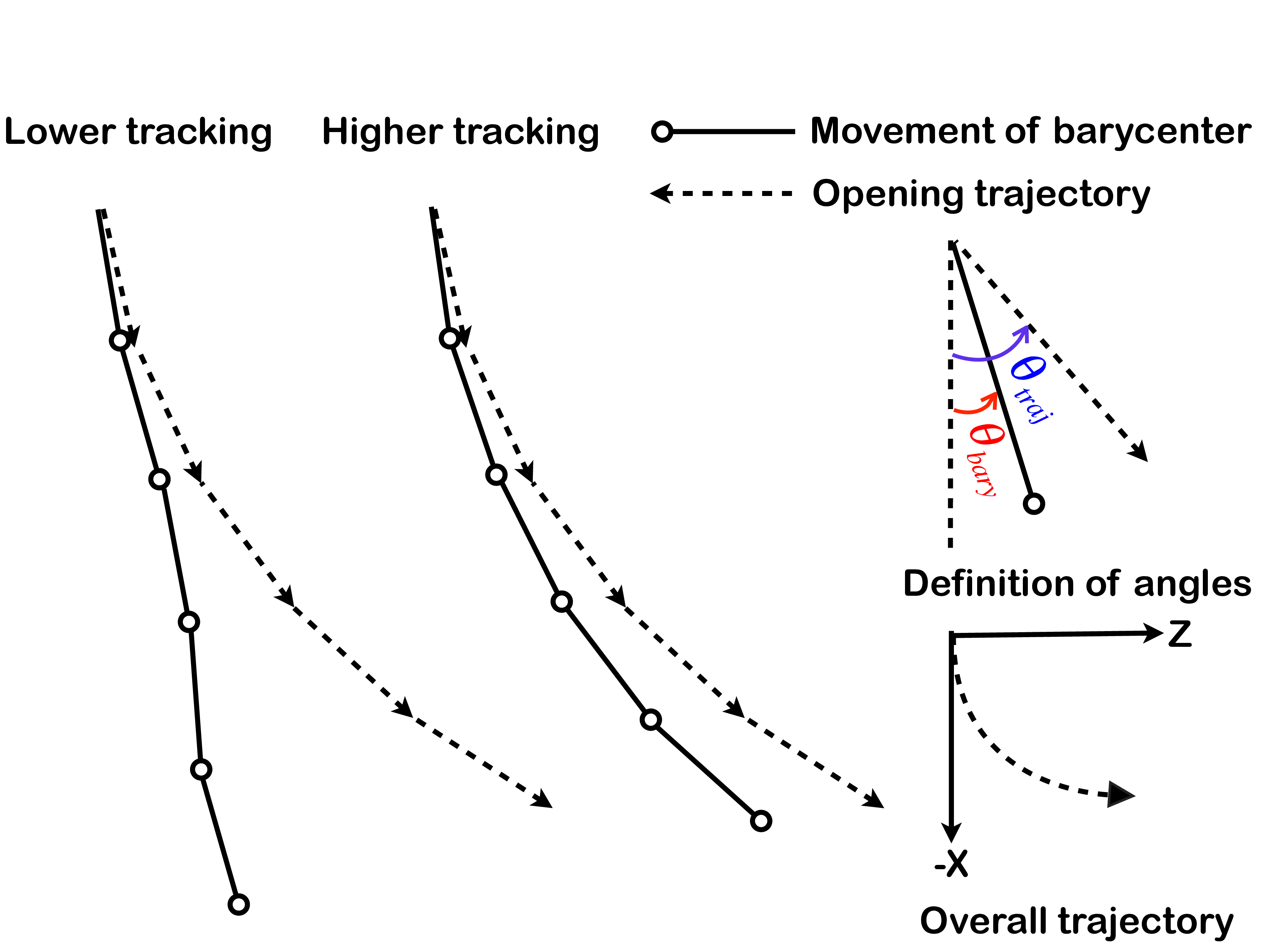}}\hspace{5 mm}
\subfigure[]{\includegraphics[scale=0.15]{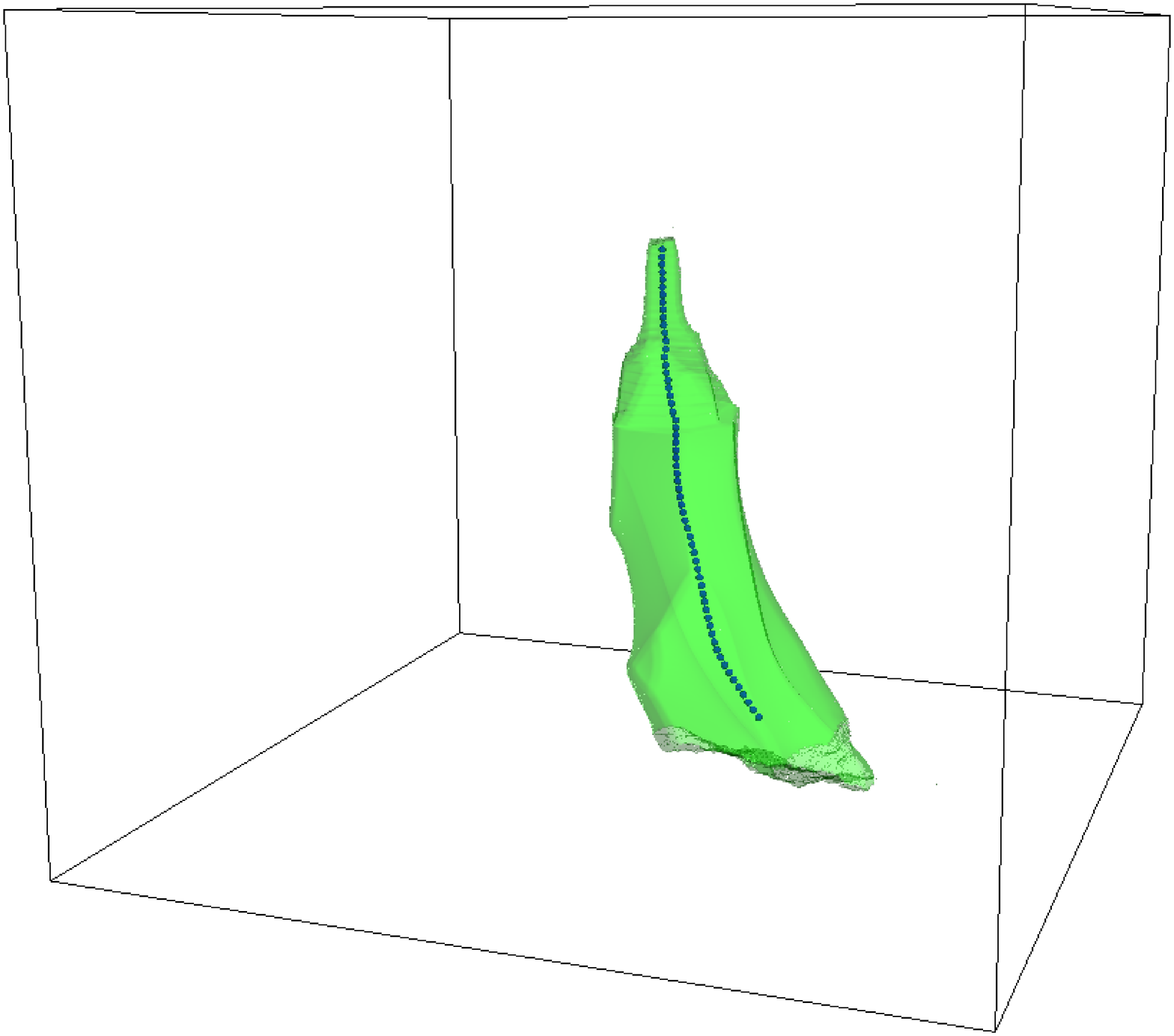}}\\
\subfigure[]{\includegraphics[scale=0.55]{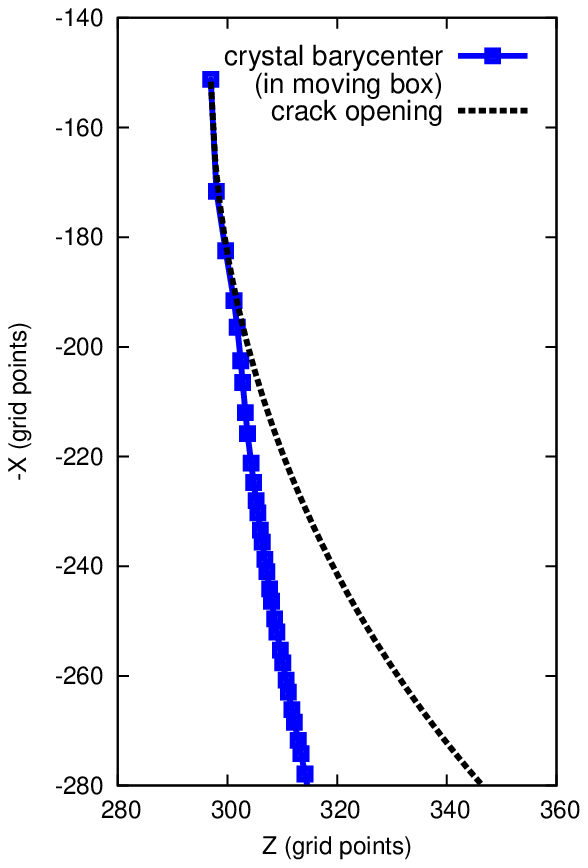}}\hspace{5 mm}
\subfigure[]{\includegraphics[scale=0.55]{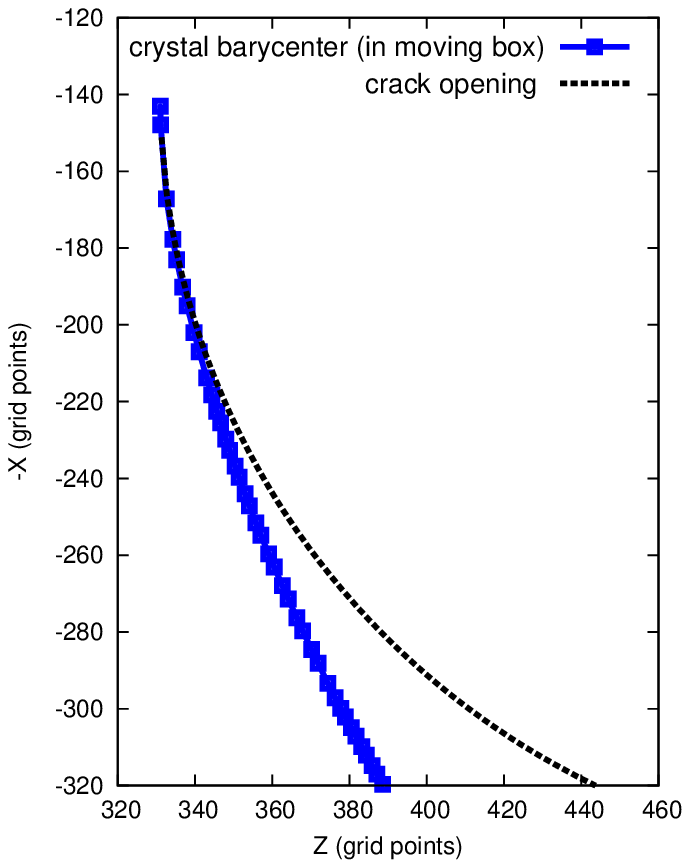}}\hspace{5 mm}
\subfigure[]{\includegraphics[scale=0.55]{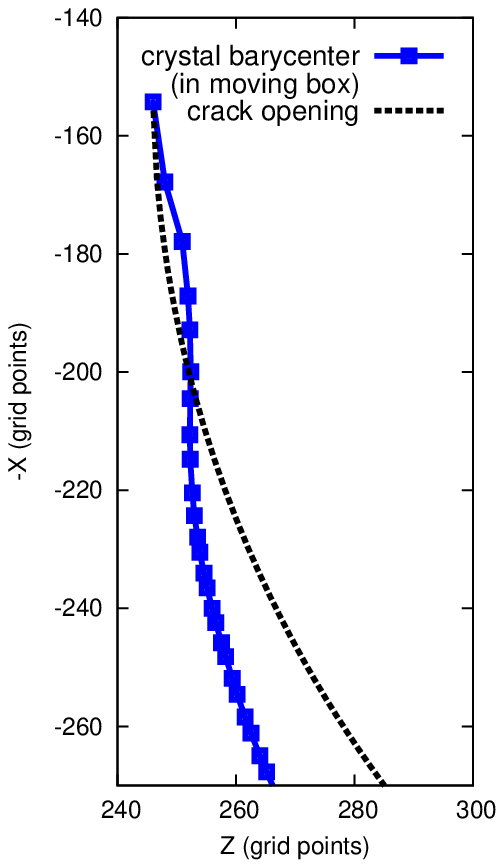}}\\
\subfigure[]{\includegraphics[scale=0.55]{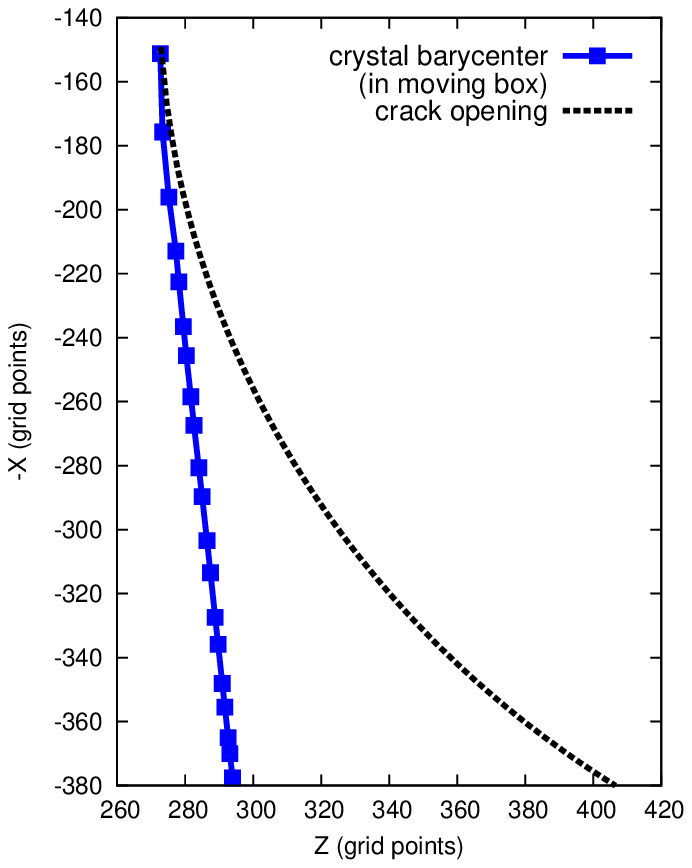}}\hspace{5 mm}
\subfigure[]{\includegraphics[scale=0.55]{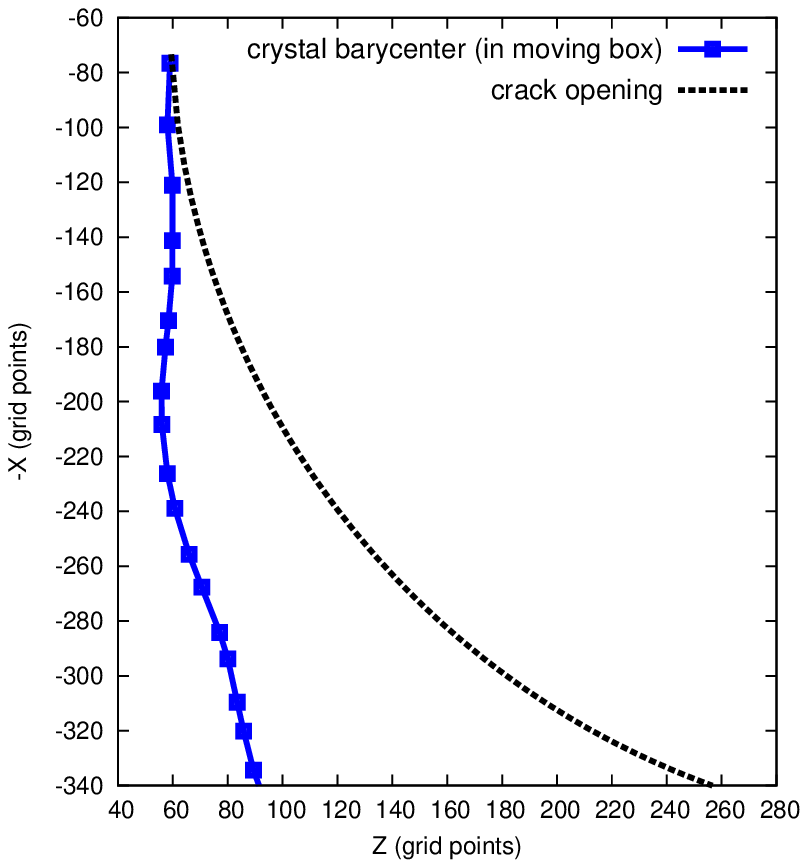}}\hspace{5 mm}
\subfigure[]{\includegraphics[scale=0.55]{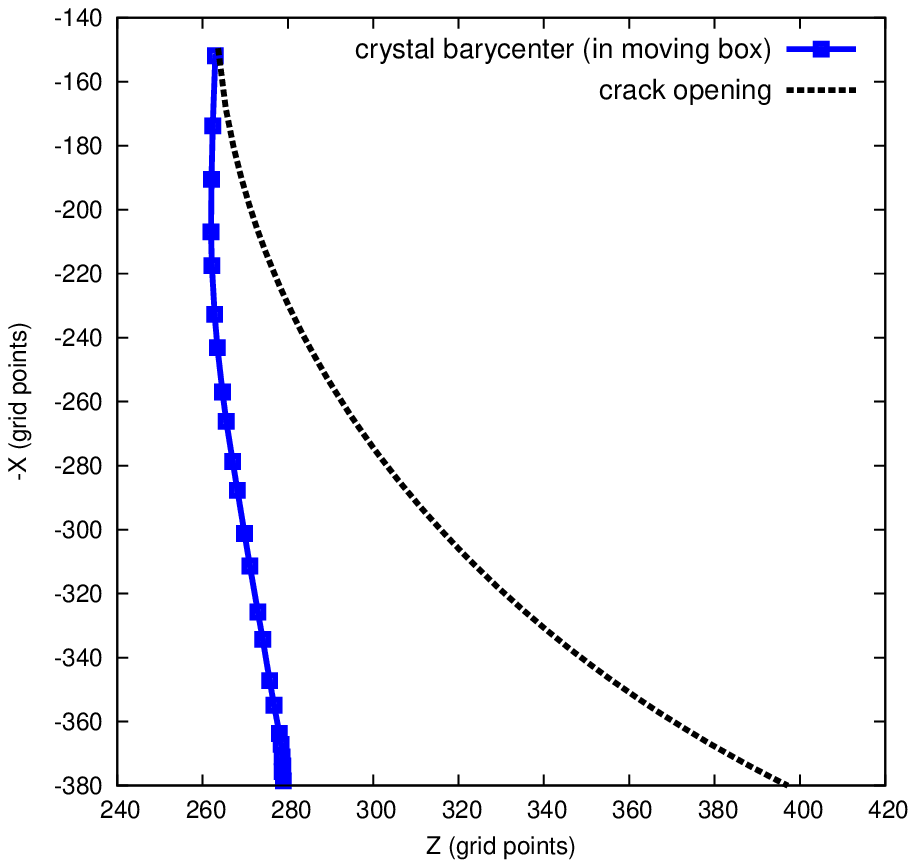}}
\caption{(a) Diagram explaining the 
calculation of GTE$_{2}$ which accounts for 
temporal evolution of tracking efficiency. The solid line 
represent the evolution of the crystal barycenter in the shifting box
while the dashed line corresponds to crack-opening trajectory.
(b) In order to account for the time evolution in calculation
of general tracking efficiency, the 
barycenter of surviving 
crystals (in contact with advancing crack surface) is determined.
For the sake of better visualization, the barycenter of one of the
surviving crystal is numerically masked over its iso-surface. 
The GTE$_2$ is calculated for surviving crystals
with axial tilts of (c) $5.63\degree$ (d) $42.45\degree$ and (e) $58.50\degree$
in simulation A  and (f) $5.63\degree$ (g) $7.16\degree$ and (h) $32.01\degree$ 
in simulation B.\label{fig:TE_bary}} 
\end{figure}

The term 'tracking efficiency' which was 
first introduced
by \citet{Urai:1991ec} is frequently 
used to quantify the tracking behavior of
crack sealing microstructures.
\citet{Ankit:2013qf} amend this definition to a  
'general tracking efficiency' 
(GTE) based on a 
fitting procedure for linear crack opening trajectory.
It is to be noted that both the 
above definitions of grain 
boundary tracking efficiency
are based on the final microstructure
morphology and do not account for
the temporal evolution of grain boundary 
tracking behavior. The dynamics
of tracking efficiency is particularly
important when the wall roughness
of the advancing crack surface 
is not sufficiently high and 
the opening
trajectory is non-linear.
For such cases, general tracking efficiency
is numerically obtained by fitting a 
straight lines in infinitesimally
small time interval $\delta t$
in which both the crack opening as well
as grain boundary morphology can be
assumed to be linear.
In the following section,
we highlight the advantage of
calculating tracking efficiency by
accounting for time evolution of grain
barycenter.
We calculate the tracking efficiency of the 
3-D computational microstructure by two different methods:
\begin{enumerate}
  \item The iso-surface of surviving crystals 
  which are in contact with the advancing wall rock
  is visualized and the local peaks (10 nearest
  neighbors approximation) 
  of the facing fractal surface are plotted.
  The peaks of the fractal surface lying along the
  grain boundaries/multi-junctions are extracted from the
  computational microstructure of the
  simulations A and B and are then superimposed, 
  as shown in figures \ref{fig:tracking_overlay_fib} and
  \ref{fig:tracking_overlay_blocky},
  respectively.
  The total number of the grain boundary/multi-junction
  tracking peaks are designated
  as $N_{tp}$.
  Similarly, those fractal peaks which neither lie 
  along grain boundaries nor at multi-junctions are plotted in 
  figures \ref{fig:non_tracking_overlay_fib} and
  \ref{fig:non_tracking_overlay_blocky}. 
  The total number of such non-tracking peaks
  is denoted by $N_{ntp}$.
  The general tracking efficiency (GTE) for the 
  3-D crack-seal microstructure
  is defined for the $i^{th}$ simulation time-step as:
  \begin{equation}
  \mathrm{GTE}_{1}^{i} = \dfrac{N_{tp}^{i}}{N_{ntp}^{i}}.
  \label{eq:GTE_1_frame}\end{equation}
  The definition of GTE$_{1}^{i}$ 
  is extended to account for 
  the overall temporal evolution of tracking efficiency
  as:
\begin{eqnarray}
\text{GTE}_{1} = \dfrac{\displaystyle\int_{0}^t
\dfrac{N_{tp}^{i}\left(t\right)}{N_{ntp}^{i}\left(t\right)}\cdot dt}{t}
= \dfrac{\displaystyle\sum_{i=1}^{n\left(\delta t\right)}
\dfrac{N_{tp}^{i}}{N_{ntp}^{i}}\cdot\left(\delta t\right)}{n\left(\delta t\right)}
= \dfrac{\displaystyle\sum_{i=1}^{n\left(\delta t\right)}
\dfrac{N_{tp}^{i}}{N_{ntp}^{i}}}{n}
\label{eq:GTE_1_temporal}
\end{eqnarray}
where $\delta t$, $n$ and $t$ represent
time-step width, number of time-steps
and total simulation time respectively.
  \begin{figure}[!htbp]
  \centering
  \subfigure[]{\includegraphics[scale=0.24]{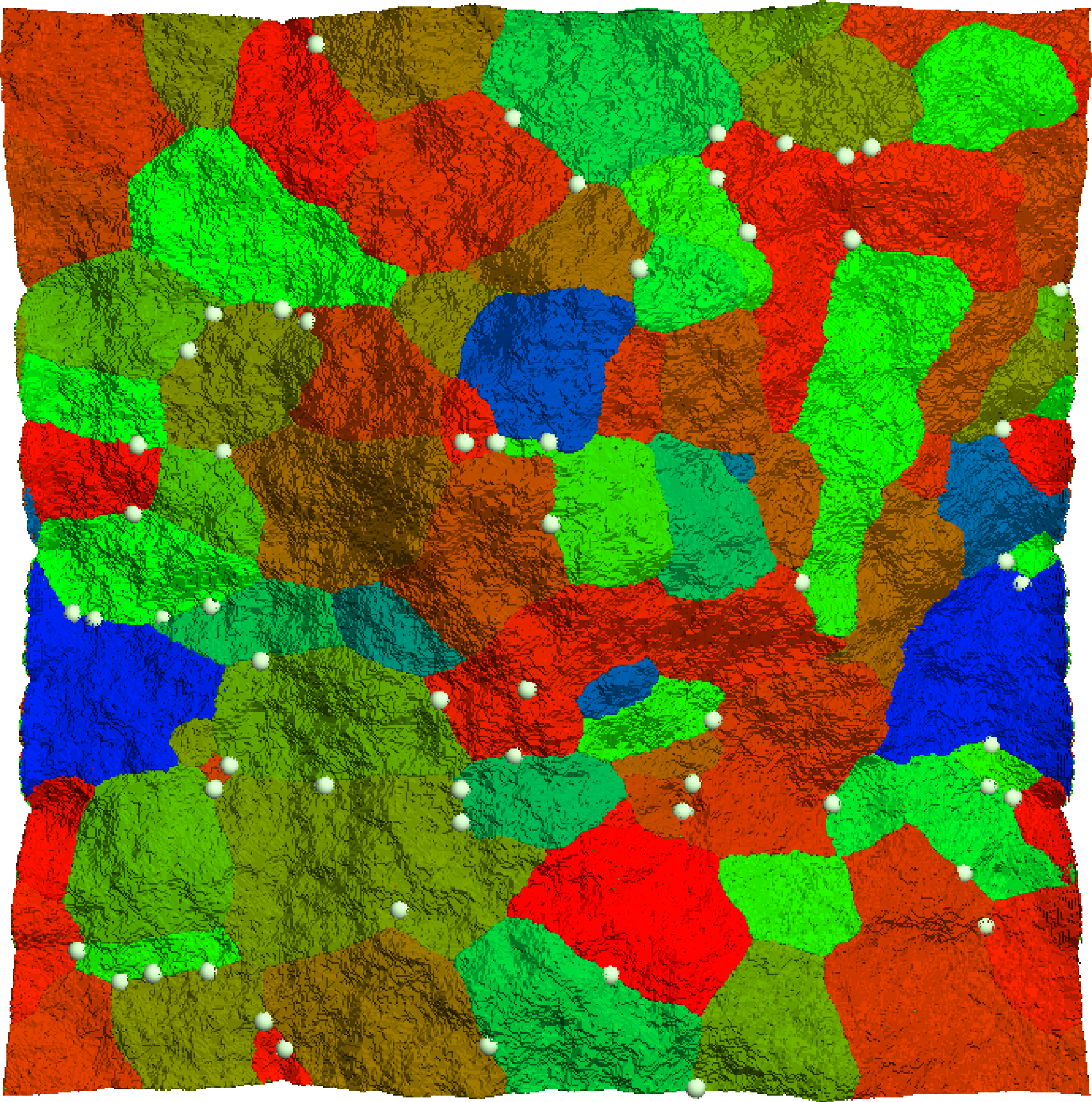}\label{fig:tracking_overlay_fib}}\qquad
  \subfigure[]{\includegraphics[scale=0.20]{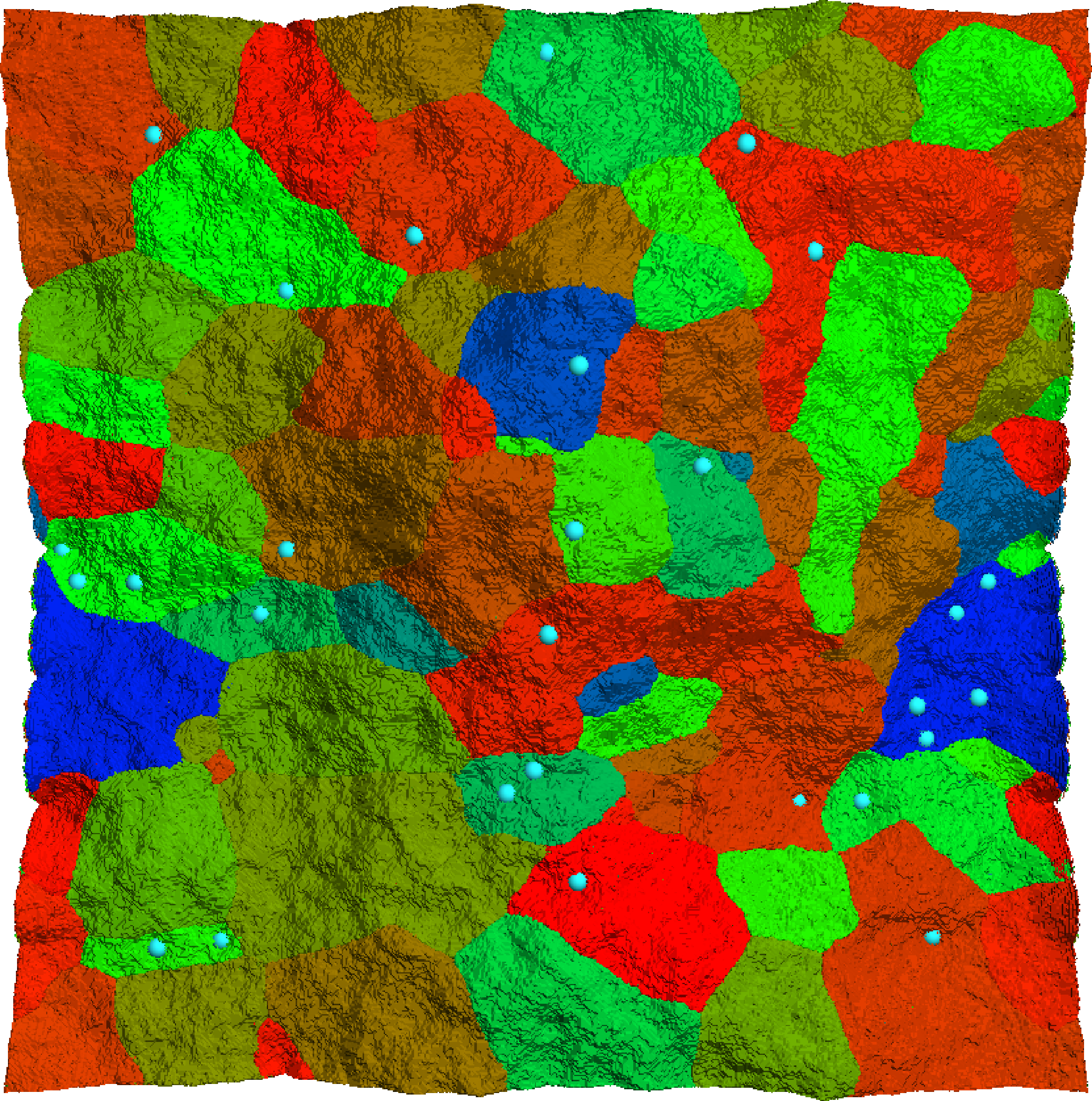}\label{fig:non_tracking_overlay_fib}}\\
    \subfigure[]{\includegraphics[scale=0.24]{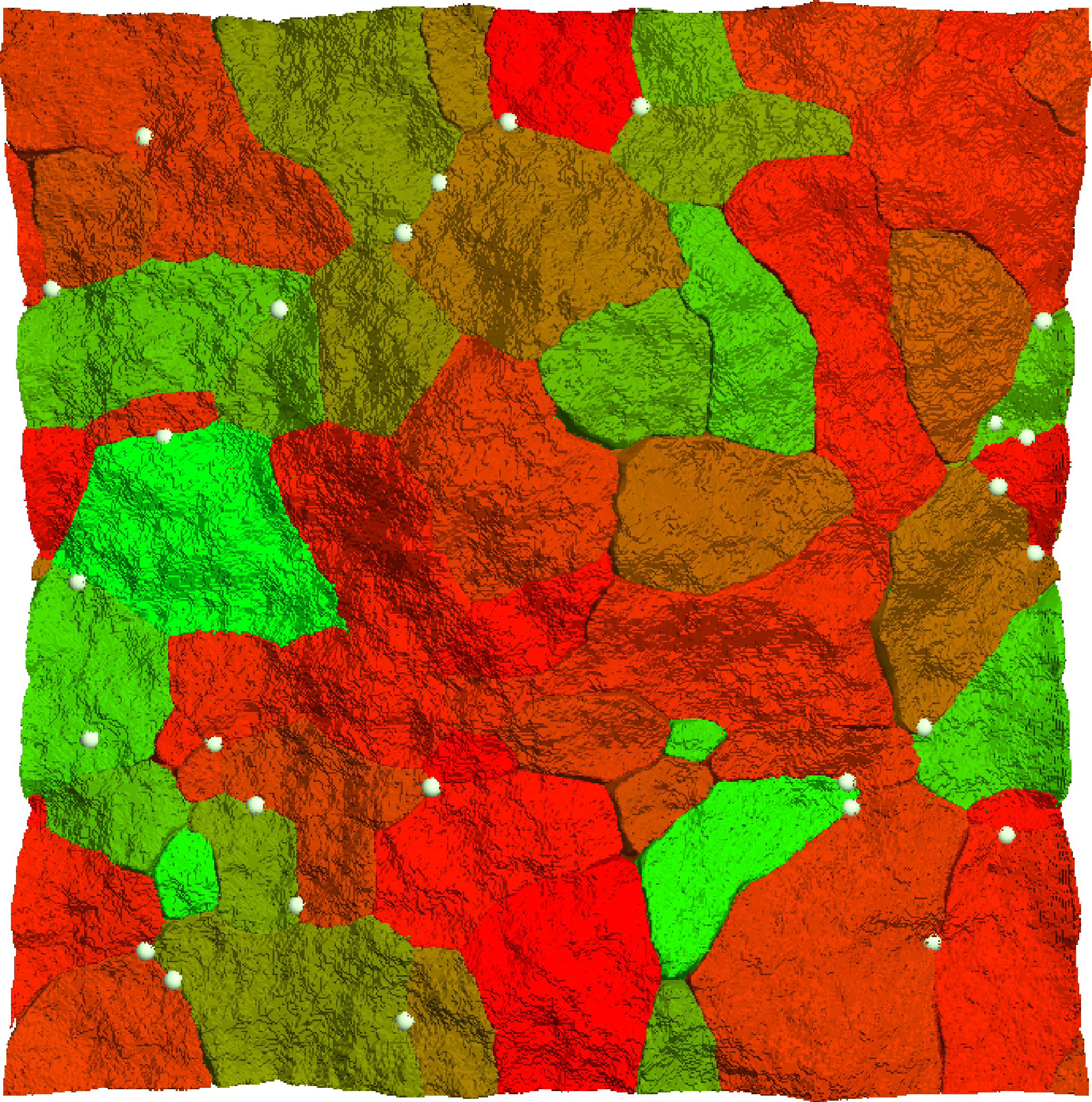}\label{fig:tracking_overlay_blocky}}\qquad
    \subfigure[]{\includegraphics[scale=0.20]{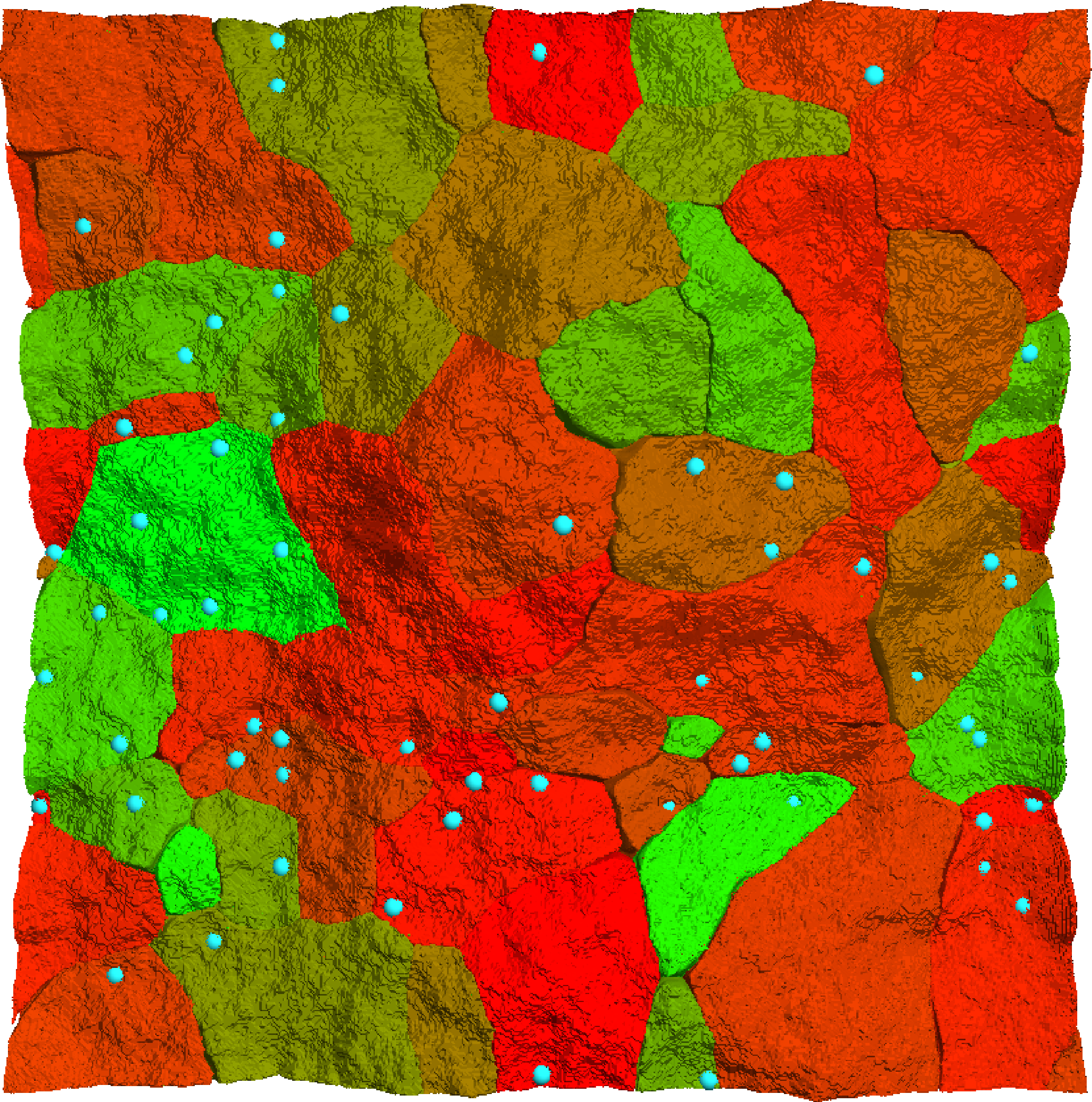}\label{fig:non_tracking_overlay_blocky}}\\
    \subfigure{\includegraphics[scale=0.5]{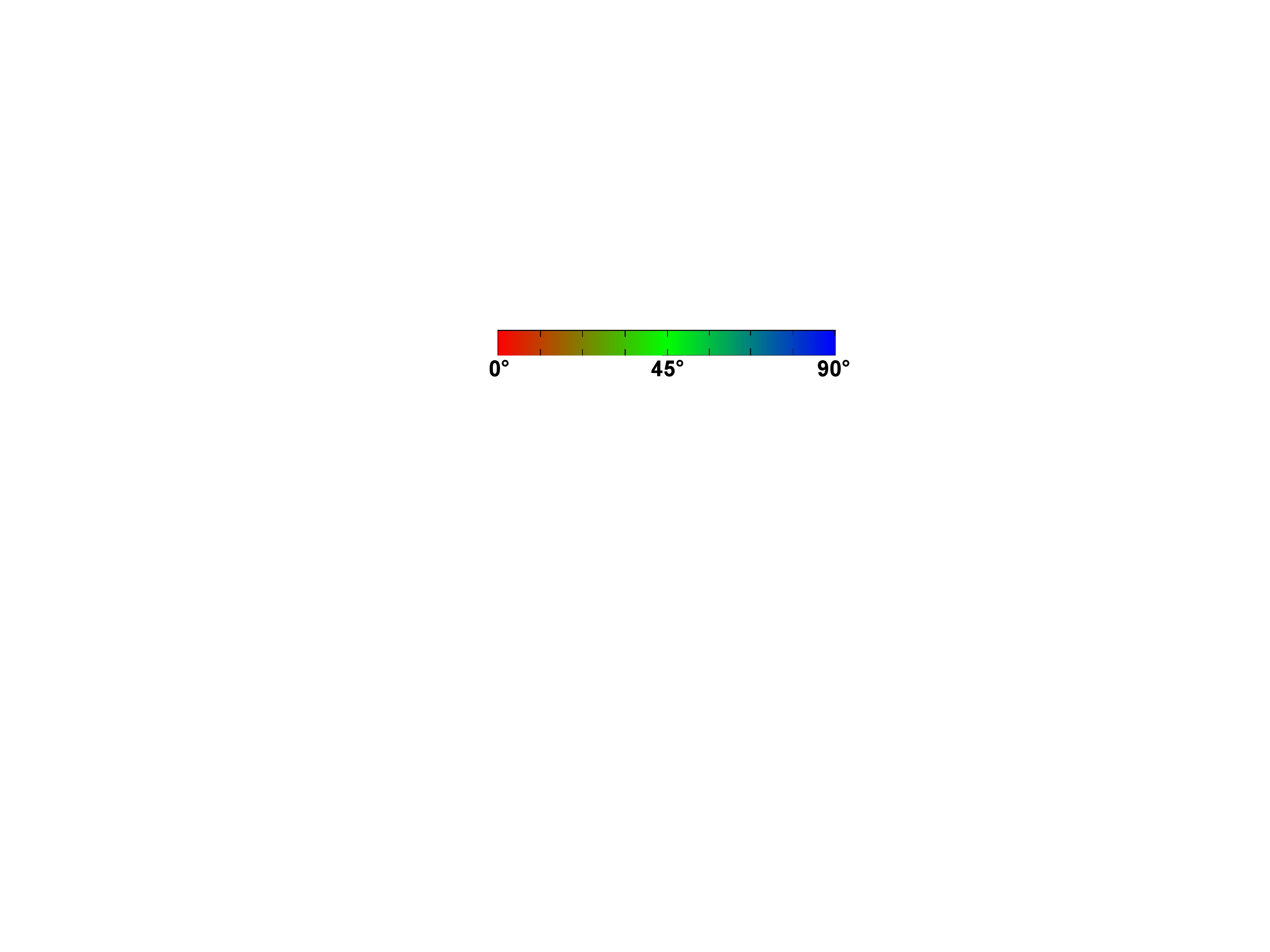}\label{fig:colorbar}}\qquad\qquad
    \subfigure{\includegraphics[scale=0.25]{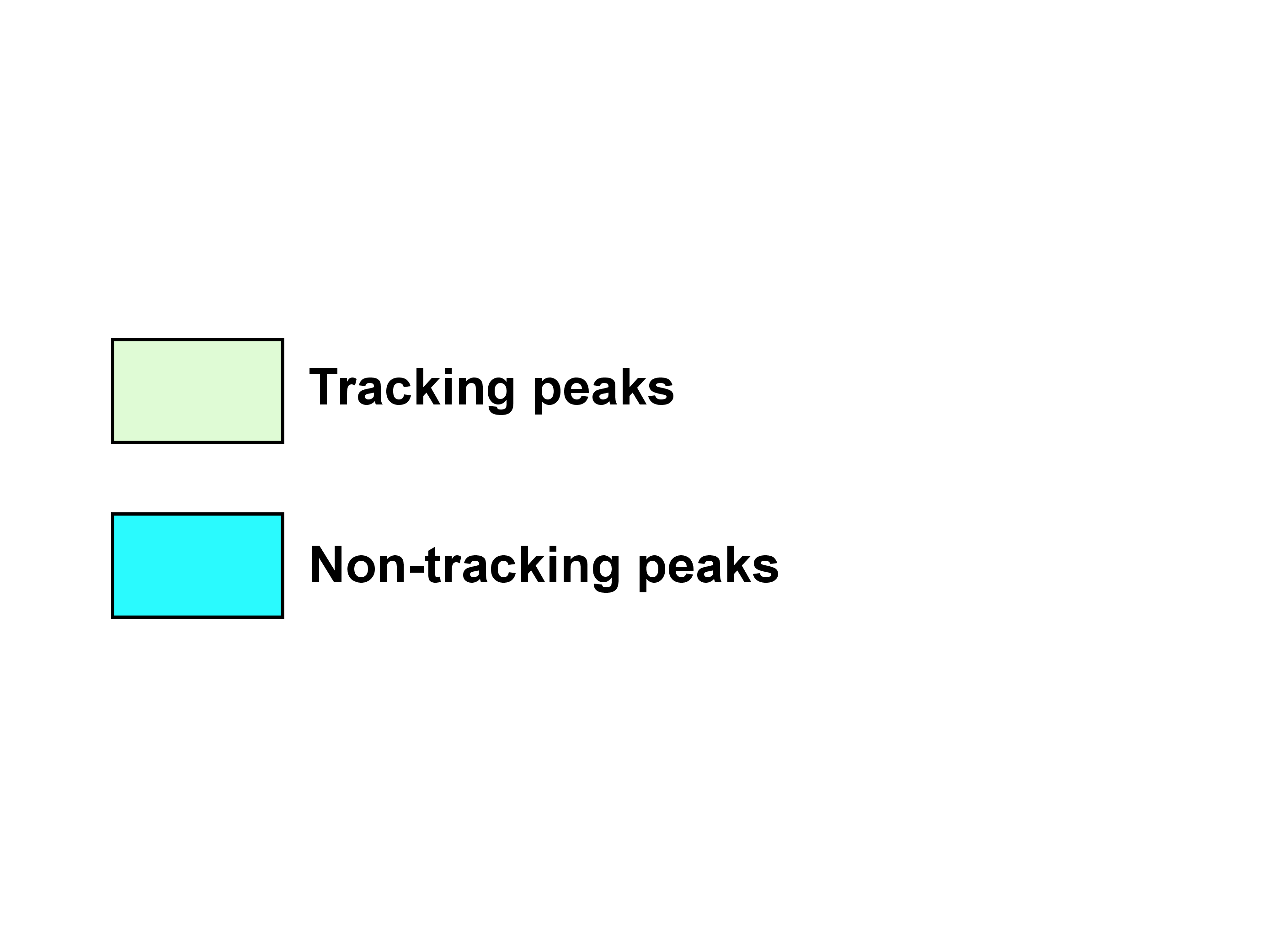}\label{fig:peak_index}}
  \caption{Local peaks of crack surface (represented as colored spheres) plotted over the rock-crystal growth interface
  for simulation A in (a) and (b) and for simulation B as shown in (c) and (d). 
  The fractal peaks tracking the grain boundaries/triple/quadruple junctions 
  are plotted as off-white spheres in (a) and (c). The fractal peaks not tracked by 
  grain boundaries/multi-junctions
  are plotted as light-blue spheres in (b) and (d).
On the basis of the final microstructures, 
the values of GTE$_{1}^{t}$ (`$t$' being 
total time) for simulation A and B are 0.685 and
0.325 respectively. On accounting for temporal evolution,
the corresponding values depreciate to 0.491 and 0.206.
The grain colors refer to 
the axial tilt indexed in the color-bar.} \label{fig:overlay}
  \end{figure}

  \item The second definition of 
  grain boundary tracking efficiency
  takes into account the temporal evolution
  of barycenter of the surviving crystals 
  in the shifting box, as described in the
  previous section. The actual positions 
  of the surviving crystal barycenters are obtained
  by adding the co-ordinates of the barycenters in the
  shifting box with the total shift
  of the simulation box as shown in figure \ref{fig:TE_bary} 
  for three tracking veins.
  For calculating the general tracking efficiency,
  the expression proposed by \citet{Ankit:2013qf} is redefined
  for a small time interval $\delta t$
  and averaged over the total simulation time, according to
  \begin{eqnarray}
\text{GTE}_{2}^{i} = \dfrac{\displaystyle\sum_{i=1}^{n\left(\delta t\right)}
\dfrac{\theta_{bary}^{i}}{\theta_{traj}^{i}}\cdot\left(\delta t\right)}{n\left(\delta t\right)}
  \end{eqnarray}
If $\delta t$ is small, we rewrite the equation to
   \begin{eqnarray}
  \text{GTE}_{2} = \dfrac{\displaystyle\int_{0}^t
  \dfrac{\theta_{bary}\left(t\right)}{\theta_{traj}\left(t\right)}\cdot dt}{t}
  \label{eq:GTE_2_temporal}
    \end{eqnarray}
\end{enumerate}


\begin{table}
\caption{Calculated general tracking efficiencies for 
the simulation test cases, A and B.
GTE$_{2}$ are comparable to the corresponding 
GTE$_{1}$ whereas GTE$_{2}^{t}$ and
GTE$_{2}^{t}$ show larger deviations.
This indicates that the general tracking
efficiencies compare
quite well provided temporal 
evolution is accounted for its estimation.}
\label{tab:GTE_2_table}      
\centering
\begin{doublespace}
\begin{tabular}{|c| c| c| c| c| c| c| c| c| c|}
\hline
\multicolumn{5}{|c|}{Simulation A} & \multicolumn{5}{c|}{Simulation B} \\
\hline
Axial tilt & GTE$_{2}$ & GTE$_{1}$ & GTE$_{2}^{t}$ & GTE$_{1}^{t}$ & Axial tilt & GTE$_{2}$ & GTE$_{1}$ & GTE$_{2}^{t}$ & GTE$_{1}^{t}$ \\
\hline
$5.63\degree$ & $0.473$ & & $0.234$ & & $5.63\degree$ & $0.166$ & & $0.167$ & \\
$42.45\degree$ & $0.422$ & {\begin{sideways}\parbox{3mm}{$0.491$}\end{sideways}} & $0.751$ & {\begin{sideways}\parbox{3mm}{$0.685$}\end{sideways}} & $7.16\degree$ & $0.204$ & {\begin{sideways}\parbox{3mm}{$0.206$}\end{sideways}} & $0.254$ &  {\begin{sideways}\parbox{3mm}{$0.325$}\end{sideways}} \\
$58.5\degree$ & $0.486$ &  & $0.639$ & & $32.01\degree$ & $0.196$ & & $0.131$ & \\
\hline
\end{tabular}
\end{doublespace}
\end{table}

The values of GTE$_{1}$ and GTE$_{2}$ are
calculated for the two simulations 
A and B. The resulting values 
for GTE$_{1}$ and GTE$_{2}$
are listed in
table \ref{tab:GTE_2_table}
and it can be seen that the 
associated temporal evolutions
compare well.
On the contrary,
GTE$_{1}^{t}$ and GTE$_{2}^{t}$
derived from final microstructures, i.e. last simulation
time-step differ significantly from 
GTE$_{1}$ and GTE$_{2}$ and
show larger deviations with respect 
to each other. Additionally, the
estimated GTE$_{2}$ vary with 
axial tilt and therefore, may lead
to erroneous interpretation.

\subsection{Statistics}
\begin{figure}
\centering
\subfigure[]{\includegraphics[scale=0.63]{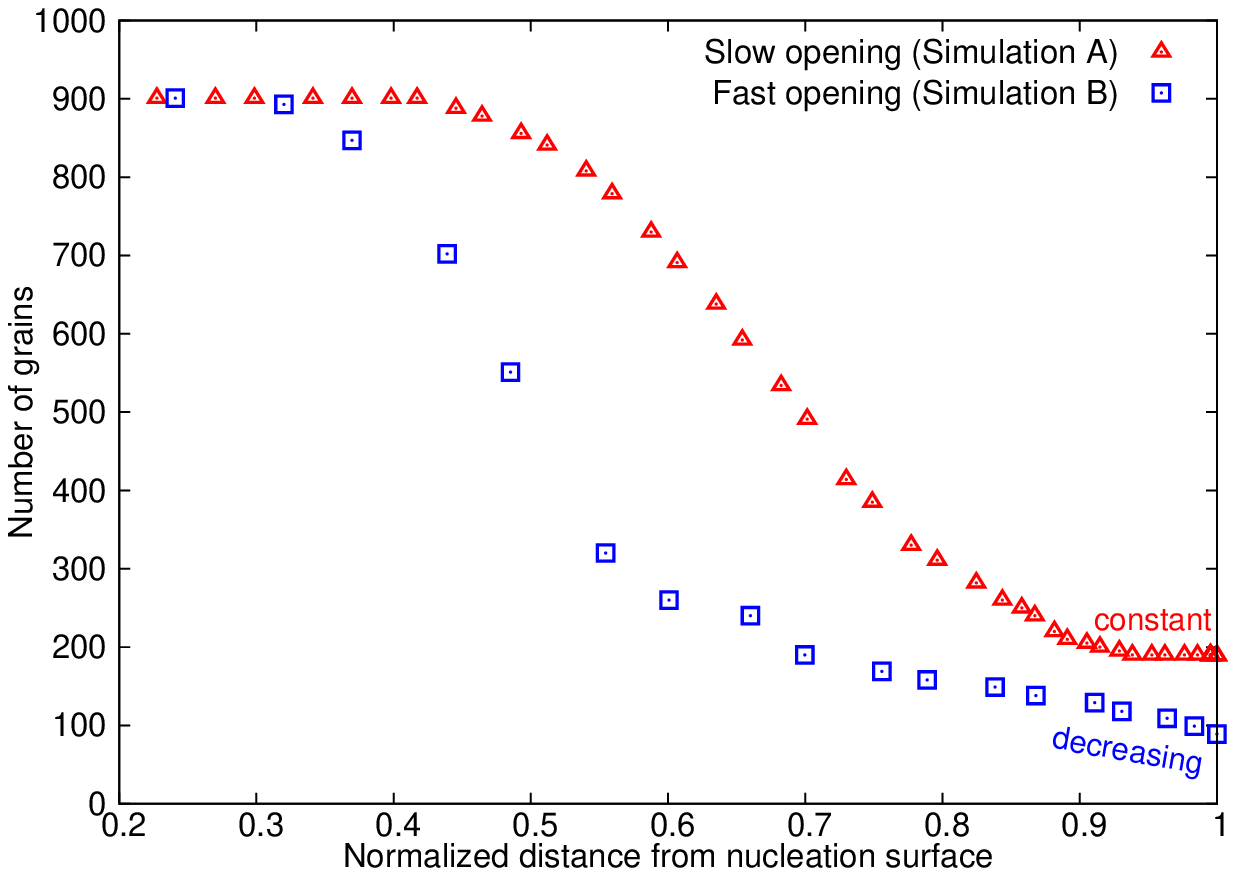}\label{fig:graincount}}\qquad
\subfigure[]{\includegraphics[scale=0.63]{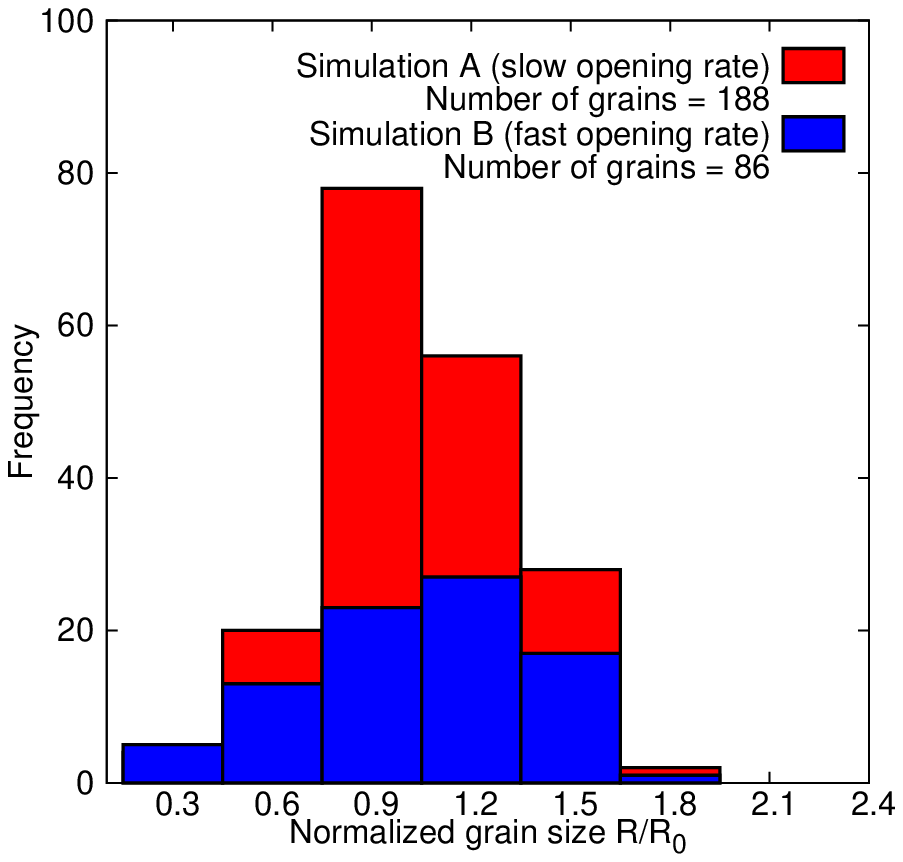}\label{fig:gsd}}
\caption{Statistics obtained from 3-D phase-field simulations. (a) Number 
of grains in contact with the advancing crack surface
plotted as function of normalized distance from the point
of greatest depression on the nucleation surface. 
The number of grains become
nearly constant when grain boundaries/triple points are
pinned at facing peaks for small crack-opening rate.
At higher crack-opening rates, the growth competition 
dominates due to a lesser `pinning effect' evident from
a decreasing trend, even at later stages.
(b) Grain size distribution for the 
final microstructures shown in figure \ref{fig:overlay}.
In the case of slower crack-opening, a shift of distribution peak 
towards smaller normalized 
grain size signifies greater pinning
leading to higher grain boundary tracking efficiency.
\label{fig:statistics}}
\end{figure}

\begin{figure}[!htbp]
\centering
\subfigure{\includegraphics[scale=0.4]{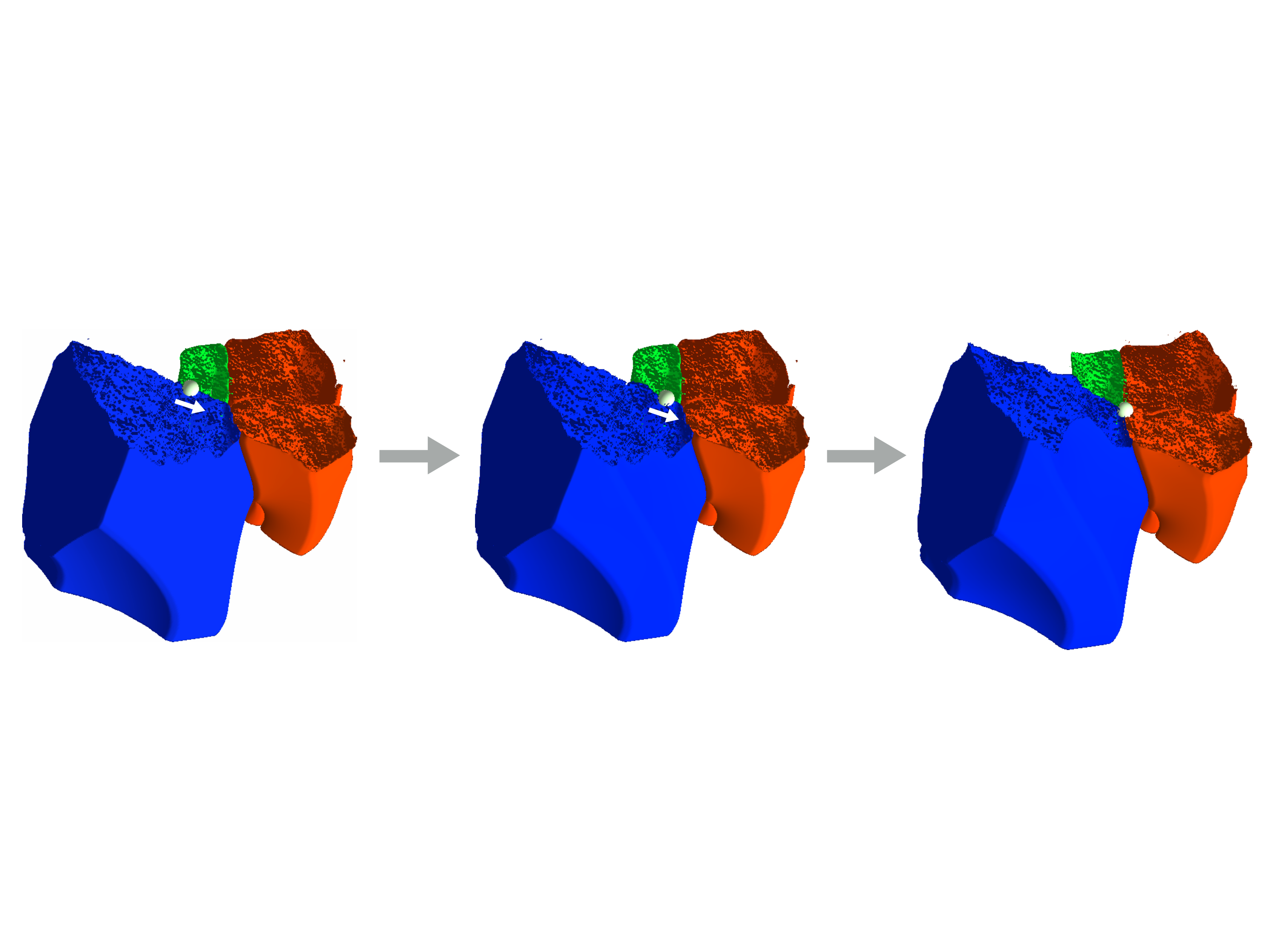}}\\
\subfigure{\includegraphics[scale=0.5]{images/colorbar.eps}\label{fig:colorbar}}
\caption{
Analysis of the 
temporal evolution 
of a fractal 
peak (in shifting box) which 
pins at the quadruple point
and results in greater 
tracking behavior, as
evident from
survival of mis-oriented
crystal fiber (in blue).
The fractal peak
illustrated as a
white sphere acts as 
a grain boundary attractor
while moving along a 
predetermined trajectory.
It is interesting to note that the 
grain quadruple
junction
acts as a stronger attractor
in comparison to grain boundaries,
in the final stages of crack-sealing
simulations (right image).
The axial tilt
of the grains are indexed
according to the colorbar.
\label{fig:quad_pinning}} 
\end{figure}

\begin{figure}[!htbp]
\centering
{\includegraphics[scale=0.4]{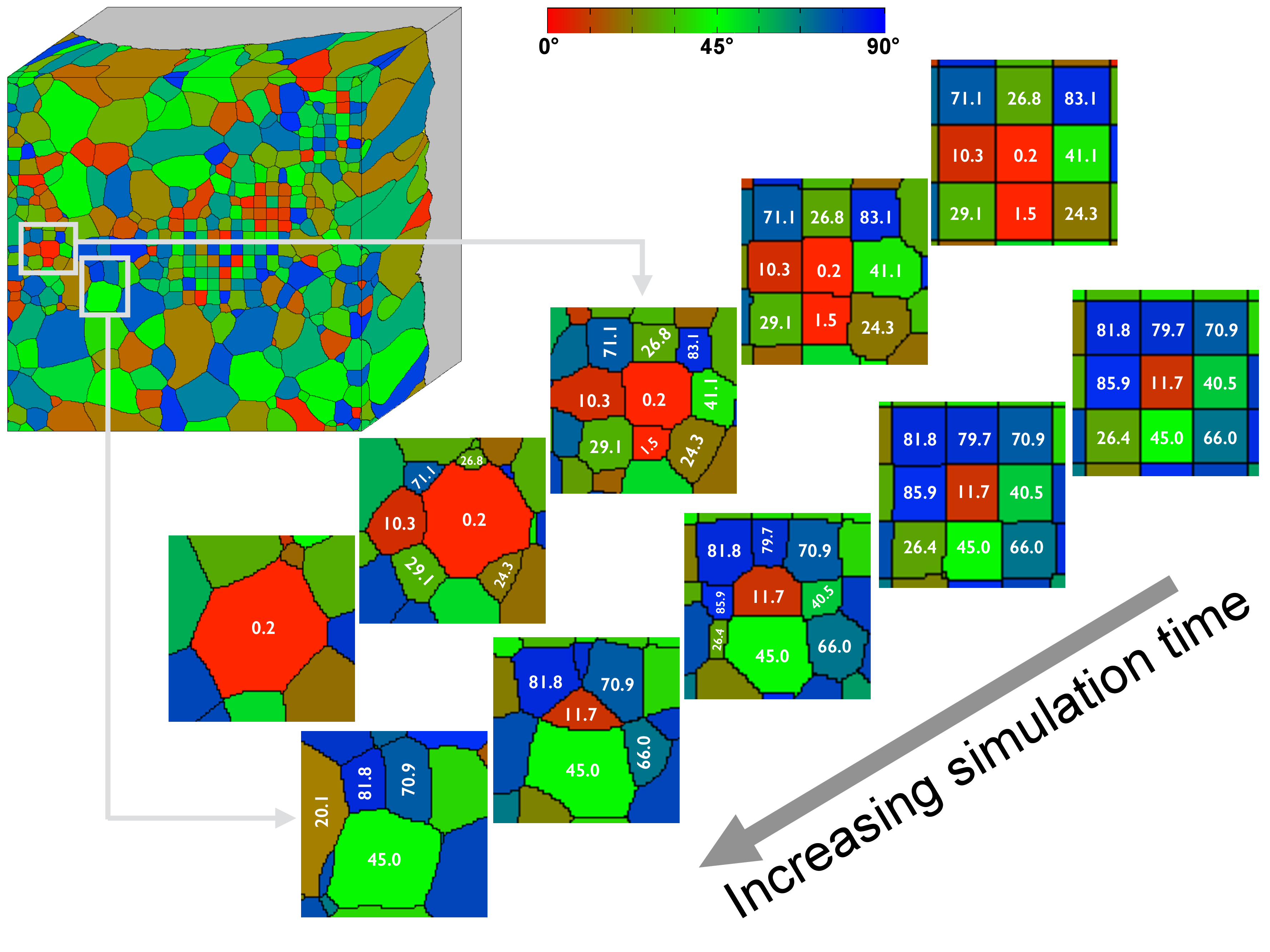}}
\caption{Temporal evolution of marked areas
in simulation A 
(slow crack-opening)
show that the
veins follow the
trajectories of the
opening peaks and
 evolve independently 
of their mis-orientation,
with respect to the 
most preferred growth direction.
In such a case, the grain boundaries/multi-junctions
whose motion is pinned by
the peaks of the advancing crack (shown in gray),
track the opening trajectory.
The color of the 
grains represent the axial tilt
(numerical values also mentioned
for grains in consideration) 
and indexed as per the colorbar.
\label{fig:grain_growth}} 
\end{figure}

It is well known from the 
2-D numerical studies
of crack-sealing
process
that a smaller 
crack-opening rate favors 
the formation of fibrous 
morphology, while, 
a larger 
crack opening rate 
leads to formation 
of blocky veins
\citep{Ankit:2013qf, Hilgers:2001rw}.
However, none of the previous 
studies focuses on the 
statistical aspects
 which have the potential
to provide valuable insights into
the vein growth process.
We plot the number 
of grains surviving the 
crack opening process
(in contact with the 
advancing crack surface) 
versus the
time dependent crack 
opening distance 
and
observe  
a shift between  
the two regimes,
in the present 3-D
simulations. As shown 
in figure \ref{fig:graincount},
the decline in the number 
of grains is significantly
steeper for
faster 
crack opening (simulation B), 
as compared to the 
case of slow opening 
(simulation A). The plummeting
of grain count is indicative of 
the anisotropy in surface
energy being dominant,
leading to orientation selection
and growth competition,
similar to free-growth conditions.
In such a case, the mis-oriented grains
are continuously eliminated by favorably
oriented neighbors. On the contrary, 
when the crack-opening rate is 
smaller, the decrease in the 
number of grains is less
steep and becomes
constant, which 
indicates that grain boundaries
are pinned by fractal peaks, 
even though the general 
tracking efficiency is 
near about 0.5 (much lesser
than 1.0).
The grain size distribution
in the shifting box (final simulation time-step) 
is plotted 
in figure \ref{fig:gsd} and
represents those grains in contact with
the advancing crack surface 
for the test cases A and B. 
On comparison,
it becomes clear that the tip
of the distribution shifts towards
smaller mean grain sizes
due to an increased pinning behavior 
of fractal peaks when the crack-opening rate
is smaller. It is noteworthy that a major 
repercussion of grain-boundary/multi-junction 
pinning as shown in figure \ref{fig:quad_pinning}
is that 
the growth competition based
on mis-orientation is suppressed. 
In such cases, the consumption
or survival of a grain does not 
depend on mis-orientation
as shown in 
figure \ref{fig:grain_growth}. 
Therefore, a higher number of 
grains survive the crack-sealing
process 
as opposed to other 
case when crack-opening
rate is faster in evidence
with the result in 
figure \ref{fig:graincount}.

\section{Discussion of results}
The present work highlights the
advantage of 3-D numerical studies
over 2-D, by visualizing the
complex and the inherently 
three-dimensional
motion of 
grain boundaries/multi-junctions 
in crack-sealing process.
In order to account 
for the temporal evolution of veins, 
two different approaches to 
determine the general 
tracking efficiencies are proposed
and compared.
We clarify the main reason
for the effect of the crack
opening rate on the
grain growth statistics,
i.e. on the 
number of grains
tracking the opening trajectory, 
size and
orientation distribution.

For implementing a
`realistic' 3-D fractal surface as
the boundary condition and obtaining a
uniform overlay of nuclei over the crack surface
which prevents the onset of size-effects,
preprocessing
algorithms are proposed.
An innovative approach
to visualize the 
numerically simulated veins
aided by post-processing techniques
reveals that the 
grain boundary/multi-junction
morphologies
in 3-D are more complicated 
as compared to 2-D cases, in general.
In order to 
deal with the
third dimensionality
which makes the
determination of
tracking efficiency difficult,
we amend the definition
of general tracking efficiency
to GTE$_{1}$ and GTE$_{2}$
given by equations 
\ref{eq:GTE_1_temporal}
and \ref{eq:GTE_2_temporal}
respectively.
While GTE$_{1}$ may be interpreted 
as an extension of the
tracking efficiency
proposed by \citet{Urai:1991ec} 
for the case of 3-D evolution,
 GTE$_{2}$ accounts
 for the temporal change in tracking
 behavior of veins
 with respect to the crack-opening trajectory.
In contrast, GTE$^{t}_{1}$ describes
the tracking efficiency of
the final microstructure (`t' being the total simulation time)
and does not account for temporal evolution.
The study of peak trajectories
demonstrates that
GTE$^{t}_{1}$ 
is  significantly
higher than the GTE$_{1}$ for
the simulation test cases, A and B.
It is noteworthy that
GTE$_{1}$ and GTE$_{2}$,
both accounting for the 
temporal evolution of 
grain boundaries/multi-junctions,
compare quite well as
summarized in table \ref{tab:GTE_2_table}.
The incongruousness 
arising out of the
neglect of temporal
evolution of grain
boundaries/multi-junctions
implies that conclusions
about the tracking efficiency, 
as previously drawn upon 
the GTE$^{t}_{1}$ data, which solely rely 
on final microstructure
have a clearly limited
validity. While the 
original approach
of using grain boundary morphology to  
determine tracking efficiency
proves to be inept in 3-D, 
we conclude that 
the proposed GTEs, 
both accounting the temporal evolution, 
serve as reliable methods to quantify
the tracking behavior in veins.

The grain evolution 
statistics obtained by post-processing
the 3-D computational microstructure 
reveals that the number of grains 
in contact 
with advancing crack surface
(figure \ref{fig:graincount}) 
decreases steeply 
when the crack-opening 
rate is higher. It can be argued
that a steep decline in the
number of grains and survival of
crystals oriented along the
most preferred orientation,
is indicative of growth
competition which does not 
relate to the boundary conditions
namely the crack surface 
roughness or the opening rate.
The near incapability
of fractal peaks to pin
the grain boundaries/multi-junctions
when crack-opening rate is higher
can be seen 
in figure \ref{fig:tracking_overlay_blocky}
and \ref{fig:non_tracking_overlay_blocky},
where reddish grains are found in majority.
On the contrary, the decline 
in the number of grains
is less steep for smaller crack 
opening rate, which finally
becomes constant. Such a 
conclusion on statistics 
can be attributed to 
stronger pinning of fractal 
peaks at grain boundaries
and multi-junctions,
which suppresses the 
growth competition based on
mis-orientation of neighboring
grains (figure  \ref{fig:grain_growth}).
An interesting outcome 
of visualizing the temporal 
evolution in figure 
\ref{fig:quad_pinning},
is that the fractal peaks
pin more strongly at 
quadruple junctions 
(grain multi-junctions
in general)
as compared to grain boundaries.
The increase in pinning behavior 
of fractal peaks at smaller
crack-opening rates is further
accentuated by shift in the
apex of grain size distribution towards 
small grain size
(figure \ref{fig:gsd}).
Thus, the plots of number of surviving grains
as well as grain size distribution 
in figure \ref{fig:statistics}
provide a statistical realization 
of the shift in regime and 
explain the vein
characteristics observed in figure
\ref{fig:grain_growth}, which is 
primarily governed by crack-opening
velocity in the present test cases.

Finally, it is worth clarifying 
that the present definitions of GTEs
are formulated with an intention 
to indicate the importance
of temporal evolution in determining
the tracking efficiency. We do not
rule out the possibility of
other definitions, which
may be equally capable to quantify 
tracking behavior in 3-D
with high precision.
Within the scope of current  
work, the prime motive is
to highlight the gain in accuracy by 
accounting for temporal evolution
in the well known 
methodologies, which strongly
emphasizes the importance of 3-D 
numerical studies. 
The growth statistics, obtained
by accounting for 
large number of grains
and supplemented by 
3-D visualization, 
aims to bridge the gap
between field observations (of 3-D layers)
and computational studies (limited 
to 2-D till date) 
and advances the
understanding of the vein growth process.

\section{Conclusion and outlook}

The current work
which is based on a 
thermodynamically consistent
approach, namely the 
phase-field method
aims to fill-in the
short-comings while focusing on
the aspects that have 
not been addressed
in the previous numerical
studies, chiefly the
large-scale 
statistical studies
as well as complex motion
of grain boundaries/multi-junctions
in 3-D.
Wherever adequate, the benefits
of 3-D simulations
have been highlighted
while enumerating the limitations
of 2-D studies.
The 3-D visualization
aided by post-processing techniques
supplement the present numerical
studies and allow to 
draw meaningful conclusions
from the simulation test cases.
The most outstanding
achievement of the present 
work is the characterization
of vein  microstructure based
on temporal evolution of 
grain boundaries/multi-junctions
rather than 
relying on an 
approximate reconstruction
from the final morphology,
a popular approach in the
geo-scientific community
\citep{Passchier:1996fk}. 
Bearing in mind that it is
fundamentally (essentially) 
difficult to design as well
as carry out in-situ studies in
laboratory experiments which can replicate
the process of vein evolution, 
the 3-D numerical studies
are of paramount importance as they 
can alternatively provide 
invaluable insights
into the vein growth process.
We further stress that the new
methodologies based on 
multiphase-field
modeling allow an efficient use 
of modern high super-computing
power, so that even the consideration
of large grain systems (up to 500,000
grains) in 3-D computational studies
becomes feasible.

\bibliographystyle{apalike}
\bibliography{Bibliography}

\begin{thebibliography}{}

\bibitem[Ankit et~al., 2013a]{Ankit:2013fk}
Ankit, K., Choudhury, A., Qin, C., Schulz, S., McDaniel, M., and Nestler, B.
  (2013a).
\newblock Theoretical and numerical study of lamellar eutectoid growth
  influenced by volume diffusion.
\newblock {\em Acta Materialia}, 61(11):4245--4253.

\bibitem[Ankit et~al., 2013b]{Ankit:2013qf}
Ankit, K., Nestler, B., Selzer, M., and Reichardt, M. (2013b).
\newblock Phase-field study of grain boundary tracking behavior in crack-seal
  microstructures.
\newblock {\em Contributions to Mineralogy and Petrology}, 166(6):1709 -- 1723.

\bibitem[Bons, 2001]{Bons:2001ve}
Bons, P. (2001).
\newblock Development of crystal morphology during unitaxial growth in a
  progressively widening vein: I. the numerical model.
\newblock {\em Journal of Structural Geology}, 23(6--7):865 -- 872.

\bibitem[Bons et~al., 2008]{Bons:2008ys}
Bons, P., Koehn, D., and Jessell, M. (2008).
\newblock {\em Microdynamics Simulation}, volume 106 of {\em Lecture Notes in
  Earth Sciences}.
\newblock Springer Berlin Heidelberg.

\bibitem[Hilgers et~al., 2001]{Hilgers:2001rw}
Hilgers, C., Koehn, D., Bons, P., and Urai, J. (2001).
\newblock Development of crystal morphology during unitaxial growth in a
  progressively widening vein: Ii. numerical simulations of the evolution of
  antitaxial fibrous veins.
\newblock {\em Journal of Structural Geology}, 23(6--7):873 -- 885.

\bibitem[Hubert et~al., 2009]{HUBERT:2009fu}
Hubert, J., Emmerich, H., and Urai, J. (2009).
\newblock Modelling the evolution of vein microstructure with phase-field
  techniques -- a first look.
\newblock {\em Journal of Metamorphic Geology}, 27(7):523--530.

\bibitem[Jessell et~al., 2001]{Jessell:2001kx}
Jessell, M., Bons, P., Evans, L., Barr, T., and St{\"u}we, K. (2001).
\newblock Elle: The numerical simulation of metamorphic and deformation
  microstructures.
\newblock {\em Computers and Geosciences}, 27(1):17--30.

\bibitem[Miller, 1986]{Miller:1986qf}
Miller, G.~S. (1986).
\newblock Definition and rendering of terrain maps.
\newblock {\em Computer Graphics (ACM)}, 20(4):39--48.

\bibitem[Nestler et~al., 2005]{Nestler:2005ye}
Nestler, B., Garcke, H., and Stinner, B. (2005).
\newblock Multicomponent alloy solidification: Phase-field modeling and
  simulations.
\newblock {\em Phys. Rev. E}, 71:041609.

\bibitem[Passchier and Trouw, 1996]{Passchier:1996fk}
Passchier, C. and Trouw, R. (1996).
\newblock {\em Microtectonics}.
\newblock Springer-Verlag Berlin Heidelberg, second edition.

\bibitem[Piazolo et~al., 2010]{Piazolo:2010uq}
Piazolo, S., Jessell, M., Bons, P., Evans, L., and Becker, J. (2010).
\newblock Numerical simulations of microstructures using the elle platform: A
  modern research and teaching tool.
\newblock {\em Journal of the Geological Society of India}, 75(1):110--127.

\bibitem[Stinner et~al., 2004]{Stinner:2004uq}
Stinner, B., Nestler, B., and Garcke, H. (2004).
\newblock A diffuse interface model for alloys with multiple components and
  phases.
\newblock {\em SIAM Journal on Applied Mathematics}, 64(3):775--799.

\bibitem[Taylor and Cahn, 1998]{Taylor:1998kq}
Taylor, J.~E. and Cahn, J.~W. (1998).
\newblock Diffuse interfaces with sharp corners and facets: Phase field models
  with strongly anisotropic surfaces.
\newblock {\em Physica D: Nonlinear Phenomena}, 112(3--4):381 -- 411.

\bibitem[Urai et~al., 1991]{Urai:1991ec}
Urai, J., Williams, P., and van Roermund, H. (1991).
\newblock Kinematics of crystal growth in syntectonic fibrous veins.
\newblock {\em Journal of Structural Geology}, 13(7):823 -- 836.

\bibitem[Wendler et~al., 2009]{Wendler:2009fk}
Wendler, F., Becker, J., Nestler, B., Bons, P., and Walte, N. (2009).
\newblock Phase-field simulations of partial melts in geological materials.
\newblock {\em Computers and Geosciences}, 35(9):1907 -- 1916.

\bibitem[Zhang and Adams, 2002]{Zhang:2002fk}
Zhang, J. and Adams, J.~B. (2002).
\newblock Facet: a novel model of simulation and visualization of
  polycrystalline thin film growth.
\newblock {\em Modelling and Simulation in Materials Science and Engineering},
  10(4):381.

\end{thebibliography}

\end{document}